\documentclass[a4paper,11pt]{article}

\usepackage{setup}
\usepackage{jheppub}
\usepackage{todonotes}
\usepackage{placeins}
\usepackage{longtable}

\usepackage{cite}
\usepackage{ifthen}
\usepackage{subcaption}
\usepackage{siunitx}
\usepackage{enumitem}
\usepackage{booktabs}
\usepackage{bookmark}

\usepackage{cleveref}
\DeclareSIUnit\fb{\femto\barn}

\usepackage[compat=1.1.0]{tikz-feynman}
\tikzfeynmanset{
  every edge/.style={line width=0.6pt, line cap=round, line join=round},
  every vertex/.style={inner sep=0pt}, % no padding around vertices
}

\renewcommand{\d}[1]{\ensuremath{\operatorname{d}\!{#1}}}
\newcommand{\sqrts}{\sqrt{s}}

\def\lapprox{\lower .7ex\hbox{$\;\stackrel{\textstyle <}{\sim}\;$}}
\def\gapprox{\lower .7ex\hbox{$\;\stackrel{\textstyle >}{\sim}\;$}}
\definecolor{lightgray}{HTML}{A6A39A}
\definecolor{darkgray}{HTML}{504E48}
\definecolor{silver}{HTML}{E0DFDE}
\definecolor{brown}{HTML}{5F4541}
\definecolor{beige}{HTML}{DCCCAC}
\definecolor{green}{HTML}{345F53}
\definecolor{yellow}{HTML}{F6B65A}
\definecolor{blue}{HTML}{568BCF}
\definecolor{red}{HTML}{AE1932}
\definecolor{orange}{HTML}{D16F15}

\newcommand{\ycut}{y_{\text{cut}}}
\newcommand{\gev}{\text{GeV}}

\makeatletter
\newcommand{\myitem}[1]{%
	\item[#1]\protected@edef\@currentlabel{#1}%
}
\makeatother

\preprint{{\raggedleft%
IPPP/25/70, ZU-TH 68/25 \\
}}

\title{NNLO QCD predictions for jet observables in $ZH$ production at electron-positron colliders}
\author[a,b]{Simone Caletti,}
\author[a,c]{Aude~Gehrmann--De Ridder,}
\author[d]{Matteo~Marcoli}

\affiliation[a]{Institute for Theoretical Physics, ETH, 8093 Z{\"u}rich, Switzerland}
\affiliation[b]{Dipartimento di Fisica, Università di Torino, and INFN, Sezione di Torino, Via Pietro Giuria 1, I-10125, Torino, Italy}
\affiliation[c]{Physik-Institut, Universit{\"a}t Z{\"u}rich, 8057 Z{\"u}rich, Switzerland}
\affiliation[d]{Institute for Particle Physics Phenomenology, Department of Physics, University of Durham, Durham, DH1 3LE, UK}

\emailAdd{simone.caletti@unito.it}
\emailAdd{gehra@phys.ethz.ch}
\emailAdd{matteo.marcoli@durham.ac.uk}

\abstract{
We present precise predictions for a variety of jet observables in $ZH$ production at electron-positron colliders, with the $Z$ boson decaying leptonically and the Higgs boson decaying into two hadronic jets, up to NNLO in perturbative QCD. 
We consider a Higgs boson decaying into bottom (charm) quark pairs via Yukawa interaction and into gluons via an effective vertex in the limit of infinite top quark mass. We present results for the two decay modes separately, highlighting relevant differences in the differential distributions, and for the sum of all decay channels, including a comparison between different choices of the Durham (or $k_T$) jet resolution parameter.
}

\begin{document}
\maketitle
\flushbottom

\section{Introduction}
\label{sec:intro}

In July 2012, the ATLAS \cite{ATLAS:2012yve} and CMS \cite{CMS:2012qbp} collaborations at the LHC reported the discovery of a scalar particle with mass near 125 GeV. At the current level of experimental accuracy, the discovered particle proves to be  consistent with the Higgs boson predicted by the Standard Model, but the limited precision of some of the measurements still leaves room for sizeable deviations. 

Up to now, a large class of couplings of the Higgs boson to Standard Model particles have been determined: the coupling of the Higgs boson to the electroweak gauge bosons $Z$ and $W$, as well as to the third-generation 
of fermions (bottom and top quarks, tau leptons) have been established at the LHC, and the associated coupling strengths measured. 
Within uncertainties, these couplings are found to be consistent with the predictions of the Standard Model of particle physics. 
A detailed and comprehensive report on the physics achievements already gained or expected from analysing LHC data can be found in \cite{deFlorian:2016spz,Cepeda:2019klc}.

In-depth precision studies of the Higgs boson properties and the associated Higgs mechanism will become possible at a future electron-positron collider, such as the FCC-ee ~\cite{Abada:2019zxq} at CERN, the CEPC~\cite{CEPCStudyGroup:2018ghi}, or the ILC~\cite{ILC:2013jhg}, which all aim to operate as ``Higgs factories" for part of their physics programme. Copious production of Higgs bosons will be achieved at a centre-of-mass energy of  $240\,\gev$, which maximises the event rate per unit time~\cite{Abada:2019zxq} for the Higgsstrahlung process $e^+e^- \to ZH$. Considering FCC-ee, in these conditions and with an integrated luminosity of approximately $10$ ab$^{-1}$, the expected number of reconstructed $ZH$ final states is in the order of a few millions. This sample size will be then comparable to the number of reconstructed 
$Z$ boson decays at LEP, which enabled an impressive range of precision studies of the electroweak interactions and QCD~\cite{ALEPH:2005ab,Bethke:2004bp}. The fundamental strength of $e^+e^-$ colliders 
comes from their clean experimental environment, with an initial state consisting of fundamental point-like particles, a precisely-known centre-of-mass energy and backgrounds many orders of magnitude lower than in hadron collider environments. A future $e^+e^-$ collider will in particular enable model-independent measurements of the Higgs coupling to gauge bosons and fermions at the level of a few percent or lower. Currently unobserved hadronic Higgs decay channels, such as the decay to gluons $H\rightarrow gg$ or charm quarks $H\rightarrow c\bar{c}$, will become accessible. 
Details regarding the expected branching ratios for all Higgs decay channels at $\sqrt{s} =240~\gev$ in the $ZH$ mode foreseen for the CERN FCC-ee project can be found in~\cite{Abada:2019zxq}.

Assuming that fully hadronic decays coming from intermediate $WW^*$, $ZZ^*$ and $\tau\tau$ states can be distinguished exploiting characteristic kinematical features~\cite{Ma:2024qoa}, the dominant hadronic Higgs decay channels are $H\rightarrow b\bar{b}$ and $H\rightarrow gg$~\cite{Djouadi:1997yw}. The $H\rightarrow b\bar{b}$ decay proceeds via a Yukawa interaction 
and has been  observed by both ATLAS~\cite{Aaboud:2018zhk} and CMS~\cite{Sirunyan:2018kst} at the LHC in the $VH$ production channel with the vector boson decaying leptonically. On the theory side, significant efforts have been made to 
 make precise predictions for cross sections and differential distributions for the Higgsstrahlung process in the kinematical regions probed by the LHC experiments and associated with the $H\rightarrow b\bar{b}$ decay.
This includes in particular the calculation of QCD effects~\cite{Ferrera:2017zex,Caola:2017xuq,Gauld:2019yng} in both the production and the decay of the Higgs boson into a bottom-quark pair, observed as two b-jets in the final state. In this case, the computations of flavoured jet observables are performed for vanishing bottom quark masses and critically rely on the tagging of bottom quarks in order to isolate the candidate jets associated with the Higgs boson. 
On the other hand, 
the $H\rightarrow gg$ decay channel, which proceeds via the top-loop-induced decay to two gluons,
is more difficult to detect at a hadron collider due to the lower branching ratio and the large QCD background from both the initial and the final state.

The inclusive widths of the two dominant hadronic decay channels are known up to fourth order in the strong coupling for the $H \to b\bar{b}$ decay with kinematically massless quarks \cite{Gorishnii:1990zu,Gorishnii:1991zr,Kataev:1993be,Surguladze:1994gc,Larin:1995sq,Chetyrkin:1995pd,Chetyrkin:1996sr,Baikov:2005rw}, with exact bottom-quark mass corrections available at second-order~\cite{Bernreuther:2018ynm}, and for the $H\to gg$ channel \cite{Spira:1995rr,Chetyrkin:1997iv,Baikov:2006ch,Davies:2017xsp,Herzog:2017dtz}
in the limit of infinitely heavy top quark mass~\cite{Wilczek:1977zn,Shifman:1978zn,Inami:1982xt}. 
In both cases, first-order electroweak corrections are available \cite{Fleischer:1980ub,Bardin:1990zj,Dabelstein:1991ky,Kniehl:1991ze,Aglietti:2004nj,Degrassi:2004mx,Actis:2008ug,Aglietti:2004ki}. The interference between the two decay modes is non-vanishing only when massive quarks are considered, and the leading mass terms are known up to fourth order in the strong coupling constant~\cite{Davies:2017xsp}. A fully-differential calculation of the $H\to b\bar{b}$ decay rate at third order in the strong coupling has been performed in \cite{Mondini:2019gid,Mondini:2019vub}. Predictions for flavoured observables associated with Higgs decays into three-jets \cite{CampilloAveleira:2024fll} and for four-jet final states~\cite{Gehrmann-DeRidder:2023uld} are available at NLO QCD. Recently, NNLO-accurate predictions for differential flavour-independent observables, such as jet rates \cite{Fox:2025cuz} and event shapes \cite{Fox:2025qmp}, including contributions from both Higgs decay categories in the Higgs rest frame have been computed. All-order resummation for event shapes in hadronic Higgs decays has been performed in~\cite{Alioli:2020fzf,Gehrmann-DeRidder:2024avt,Fox:2025txz}.

Regarding the Higgsstrahlung process at  $e^+e^-$ colliders, physics studies so far focused on precise determinations of the $ZZH$ vertex and of Higgs decay branching ratios~\cite{Moortgat-Picka:2015yla,Knobbe:2023njd}. To our knowledge, differential predictions associated with hadronic Higgs decays in the $e^+e^-\to ZH$ process, have not yet been computed beyond NLO QCD. In view of prospective Higgs coupling measurements related to $ZH$ production at  future $e^+e^-$ colliders, it is of crucial 
importance to have precise theoretical predictions matching the expected experimental accuracy. It is therefore the goal of this paper to provide precise theoretical predictions for differential distributions related to the Higgsstrahlung process $e^+e^- \to ZH$ at $\sqrt{s} =240~\gev$, with the Z-boson decaying leptonically and the Higgs boson decaying hadronically into two flavourless final state jets,  including QCD corrections up to NNLO.

The computation includes both the $H\rightarrow b\bar{b}$ and $H\rightarrow gg$  decay modes, considering kinematically massless bottom quarks and an effective $Hgg$ vertex in the limit of infinite top-quark mass. In the limit of vanishing bottom quark masses used here, the two decay processes do not interfere at any order in perturbation theory and higher order corrections to both processes can be computed independently. We consider a series of jet observables: jet rates, leading and sub-leading jet energy, angle between the two leading jets, and angle between the negatively-charged lepton and the closest jet. We study the impact of higher-order QCD corrections in the two modes highlighting relevant differences, provide phenomenological predictions for the sum of all hadronic decay channels including the decay to charm quarks, and study the effects of different jet resolution parameter choices. The differential distributions exhibit characteristic features which can be straightforwardly interpreted in terms of the kinematics of the underlying electroweak $ZH$ production process. 
Compared to differential results obtained so far for Higgs decays at NNLO~\cite{Mondini:2019vub,Mondini:2019gid,Fox:2025cuz,Fox:2025qmp},
we do not boost the Higgs boson decay products to the Higgs rest frame, but investigate the full $ZH$ production process in the $e^+e^-$  laboratory frame.

The outline of the paper is as follows: 
in Section~\ref{sec:framework} we summarise our framework and present the essential ingredients entering the calculation, in Section~\ref{sec:results} we present our numerical results, and we summarise our main findings in Section~\ref{sec:conclusions}. 

%%%%%

\section{General framework and ingredients of the computation} 
\label{sec:framework}

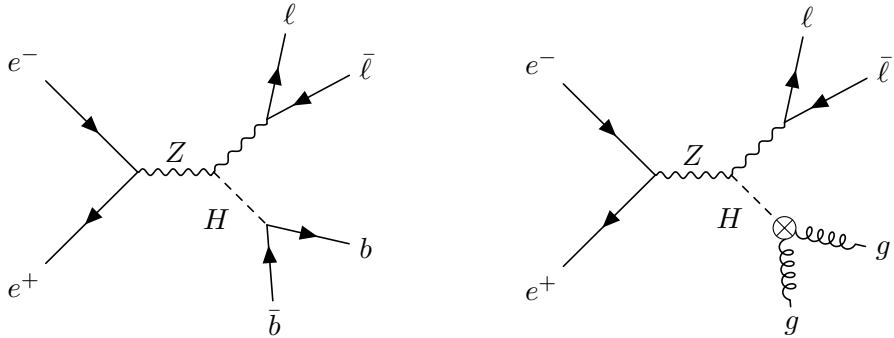
\begin{figure}
	\centering
	\begin{tikzpicture}
  \begin{feynman}
    % Incoming e+ e-
    \vertex (i1) at (-1.5,1.5) {\(e^-\)};
    \vertex (i2) at (-1.5,-1.5) {\(e^+\)};
    \vertex (v1) at (0,0);
    \vertex (v2) at (1,0);
    % Z boson and Higgs splitting
    \vertex (vz) at (1.7,0.7);
    \vertex (vh) at (1.7,-0.7);
    % Z decay into leptons
    \vertex (lz1) at (2,2.1) {\(\ell\)};
    \vertex (lz2) at (3,1.4) {\(\bar\ell\)};
    % H decay into b quarks
    \vertex (hb1) at (3,-1) {\(b\)};
    \vertex (hb2) at (1.8,-2) {\(\bar{b}\)};
    % Draw the diagram
    \diagram*{
      (i1) -- [fermion] (v1) -- [fermion] (i2),
      (v1) -- [boson, edge label=\(Z\)] (v2),
      (v2) -- [boson] (vz),
      (v2) -- [scalar, edge label'=\(H\)] (vh),
      (vz) -- [fermion] (lz1),
      (vz) -- [anti fermion] (lz2),
      (vh) -- [fermion] (hb1),
      (vh) -- [anti fermion] (hb2),
    };
  \end{feynman}
\end{tikzpicture}\hspace{1.5cm}
	\begin{tikzpicture}
  \begin{feynman}
    % Incoming e+ e-
    \vertex (i1) at (-1.5,1.5) {\(e^-\)};
    \vertex (i2) at (-1.5,-1.5) {\(e^+\)};
    \vertex (v1) at (0,0);
    \vertex (v2) at (1,0);
    % Z boson and Higgs splitting
    \vertex (vz) at (1.7,0.7);
    \node[crossed dot] (vh) at (1.7,-0.7);
    % Z decay into leptons
    \vertex (lz1) at (2,2.1) {\(\ell\)};
    \vertex (lz2) at (3,1.4) {\(\bar\ell\)};
    % H decay into gluons
    \vertex (hg1) at (3,-1) {\(g\)};
    \vertex (hg2) at (1.8,-2) {\(g\)};
    % Draw the diagram
    \diagram*{
      (i1) -- [fermion] (v1) -- [fermion] (i2),
      (v1) -- [boson, edge label=\(Z\)] (v2),
      (v2) -- [boson] (vz),
      (v2) -- [scalar, edge label'=\(H\)] (vh),
      (vz) -- [fermion] (lz1),
      (vz) -- [anti fermion] (lz2),
      (vh) -- [gluon] (hg1),
      (vh) -- [gluon] (hg2),
    };
  \end{feynman}
\end{tikzpicture}
	\caption{Feynman diagrams for the production of a $Z$ and a Higgs boson at electron-positron colliders, with leading order hadronic decays of the Higgs boson to bottom quarks via Yukawa interaction and gluons via an effective vertex denoted as a crossed dot.}
	\label{fig:diagrams}
\end{figure}

In this section, we present the main ingredients that enter the calculation of jet observables related to the Higgsstrahlung process $e^+e^-\rightarrow ZH \rightarrow \ell\bar\ell\, b\bar{b}\,(gg)$, where the vector boson $Z$ decays leptonically and the Higgs boson decays hadronically into two or more hard jets. Both the $Z$ and the Higgs boson are treated off-shell. We consider the main hadronic decay modes for the Higgs boson: to massless bottom quarks via a Yukawa-type coupling, and to gluons via an effective interaction~\cite{Wilczek:1977zn,Shifman:1978zn,Inami:1982xt} derived in the limit where the mass of the top quark circulating in the loop is infinite. 
Feynman diagrams entering the leading order contributions to the parton-level cross section associated with these two Higgs decay categories are presented in Figure~\ref{fig:diagrams}.

We include QCD corrections to the hadronic decay of the Higgs boson up to NNLO in the two decay modes. At each perturbative order, we can express the fully differential cross section in a factorised form
\begin{align}
\d\sigma^{\text{N}^k\text{LO}}_X = \sum_{i=0}^k \int \text{d} q^2 \d\sigma_{ZH} \times \frac{1}{(q^2-m_H^2)^2 + \Gamma_H^2m_H^2} \times \d\Gamma_{H\to X}^{(i)}\,,\quad\text{with}\quad X = b\bar b,\,  gg\,,
\end{align}
where $\d\sigma_{ZH}$ includes the electroweak production of a $Z$ and a Higgs boson and the leptonic decay $Z\to\ell\bar{\ell}$. The Higgs boson mass and width are respectively denoted with $m_H$ and $\Gamma_H$, while $\d\Gamma_{H\to X}^{(i)}$ indicates the differential decay rate to hadrons,
%at $i$-th order in QCD. 
with  $i$ indicating the power of the strong coupling constant $\alphas$ taken into account.
The LO inclusive Higgs decay widths for the two channels read
\begin{equation}
	\Gamma^{(0)}_{H\to b\bar{b}} = \frac{y_b^2(\muR)m_H\NC}{8\pi} \,, \quad \Gamma_{H\to gg}^{(0)} = \frac{\lambda_0^2(\muR)m_H^3(\NC^2-1)}{64\pi} \, .
	\label{eq:ratesLO}
\end{equation}
The Yukawa coupling of the $b$-quark, $y_b$, and the LO effective $Hgg$ coupling in the heavy-top limit, $\lambda_0$, are given by
\begin{equation}\label{eq:couplings}
	y_b^2(\muR) = m_b^2(\muR)\sqrt{2}\GF\,, \quad \lambda_0^2(\muR) = \frac{\alphas^2(\muR)\sqrt{2}\GF}{9\pi^2}\,,
\end{equation}
where $\GF$ is the Fermi constant. We renormalise both quantities above in the $\overline{\text{MS}}$ scheme. The running of Yukawa coupling and of the top-quark mass, which enters higher-order corrections of the $Hgg$ Wilson coefficient, is performed according to \cite{Vermaseren:1997fq}. It is important to stress that due to their different chirality structure, the two Higgs decay modes do not interfere under the assumption of kinematically massless quarks. In particular, they do not mix under renormalisation \cite{Gao:2019mlt}. This allows us to define two separate Higgs-decay categories, as mentioned above, and to compute higher-order QCD corrections independently for each decay class. 
Moreover, for massive $b$ quarks, interference contributions are sub-percent level for jet observables in hadronic Higgs decays \cite{Fox:2025cuz}, those can therefore safely be neglected.

Higher-order QCD corrections to jet observables are computed using the antenna subtraction method~\cite{Gehrmann-DeRidder:2005btv,Currie:2013vh}, relying on antenna functions extracted from physical matrix elements for the decay of a colour singlet to hadrons~\cite{Gehrmann-DeRidder:2004ttg,Gehrmann-DeRidder:2005alt}. The calculation is implemented in the NNLOJET Monte Carlo framework \cite{NNLOJET:2025rno}, suitably extended to include the considered Higgsstrahlung process $e^+e^-\rightarrow ZH \rightarrow \ell\bar\ell\, b\bar{b}\,(gg)$. 

We found agreement at lower orders between our implementation and Sherpa~\cite{Sherpa:2024mfk} for the different Higgs decay modes separately. We also verified that for the fully inclusive cross section, at each perturbative order we obtain:
\begin{equation}
	\dfrac{\sigma^{\text{N}^k\text{LO}}_X}{\sigma^{\text{LO}}_X}=\dfrac{\Gamma^{\text{N}^k\text{LO}}_{H\to X}}{\Gamma^{\text{LO}}_{H\to X}}\quad\text{with}\,\,X=b\bar{b}, \,gg,\quad\text{and}\,\,k=1,2,
\end{equation}
where $\Gamma^{\text{N}^k\text{LO}}_{H\to X}$ is the inclusive decay rate of a Higgs boson to bottom quarks or gluons at N$^k$LO in agreement with the analytic expressions presented in~\cite{Herzog:2017dtz}.

\section{Numerical results}
\label{sec:results}

We provide predictions for inclusive jet observables in $ZH$ production in $e^+e^-$ collisions at $\sqrts = 240$ GeV. 
We work in the $G_\mu$ scheme with constant electroweak parameters, given by
\begin{equation}
\GF = 1.1664 \times 10^{-5}\,\text{GeV}^{-2}\,,\quad m_Z = 91.188\,\gev\,,\quad m_H= 125.09\,\gev\,.
\end{equation}
We use $\mu_R = m_H$ as our central scale, and the perturbative uncertainty is evaluated as customary by considering predictions at $\mu_R = m_H/2$ and $\mu_R = 2 m_H$. The reference value for the strong coupling constant is $\alpha_S(m_Z) = 0.11800$ and its evolution is performed with LHAPDF6~\cite{Buckley:2014ana}, which yields $\alpha_S(m_H) = 0.11263$. The Fermi constant determines the Higgs boson vacuum expectation value $v=(\sqrt{2}\GF)^{-1/2}=246.22$ GeV.
We present the results for the Yukawa-induced decay to bottom quarks, the decay to gluons and the sum of all hadronic decay modes. For the gluonic decay mode, we rescale $\lambda_0^2(\mu_R)$ in~\eqref{eq:couplings} to include finite top, bottom and charm mass effects in the effective coupling~\cite{Spira:1997dg}, and electroweak corrections~\cite{Actis:2008ug}. In the sum, the Higgs decay to charm quarks is included as well, as it has a significant phenomenological impact on the predictions. It can be directly obtained by rescaling the results of the decay to bottom quarks by the squared ratio of the Yukawa couplings $y_c(\mu_R)^2/y_b(\mu_R)^2$. We work with running bottom and charm quark Yukawa couplings in the  $\overline{\text{MS}}$ scheme \mbox{$y_b(m_H)=m_b(m_H)/v=0.011309$} and \mbox{$y_c(m_H)=m_c(m_H)/v=0.0024629$}. Besides these Yukawa couplings, the five light quarks are treated as massless in the computation.  
We further use the $\overline{\text{MS}}$ top quark mass \mbox{$m_t(m_H) = 166.48~\GeV$}.

The rest of this section is divided into three subsections: \ref{sec:jetrates},~\ref{sec:energies} and~\ref{sec:angles}, where we discuss results for the following jet observables:  jet rates, leading- and subleading-jet energy, and angular observables respectively. For the jet rates discussed in Section~\ref{sec:jetrates}, we normalise the results with respect to the inclusive cross section at NNLO, to ensure that the rates add up to unity.
For differential distributions presented in Sections~\ref{sec:energies} and~\ref{sec:angles}, we show results normalised with respect to the LO inclusive cross section in the appropriate decay channels. Namely, distributions for the \mbox{$H\to b\bar{b}$} channel,  the $H\to gg$ channel, and the total sum of all decay channels are normalised respectively by $\sigma^0_{H\to b\bar{b}}$, $\sigma^0_{H\to gg}$ and $\sigma^0_{\text{tot.}}=\sigma^0_{H\to b\bar{b}}+\sigma^0_{H\to gg}+\sigma^0_{H\to c\bar{c}}$. Numerical values for the inclusive cross sections at LO are given in Table~\ref{tab:LOxs}.
\begin{table}[h]
	\centering
	\renewcommand{\arraystretch}{1.3} % increase cell height
	\begin{tabular}{lcccc}
		\hline
		& $H \to b\bar{b}$ & $H \to c\bar{c}$ & $H \to g g$ & Total \\
		\hline
		$\sigma^{(0)}$ &  $3.6828$ & $0.17455 $ & $0.35515$ & $4.2125$ \\
		\hline
	\end{tabular}
	\caption{Numerical values in fb for the LO cross section in different Higgs decay modes.}
	\label{tab:LOxs}
\end{table}

\subsection{Jet rates}
\label{sec:jetrates}

Final state partons are clustered into jets with the Durham (or $k_T$) algorithm \cite{Catani:1991hj}, which uses the distance measure 
\begin{align}
	y^D_{ij} = \frac{2\min[E_i^2,E_j^2]}{Q^2}(1-\cos\theta_{ij})\,,
\end{align}
where $Q$ is the total mass of the hadronic system, which for the considered process is the mass of the (off-shell) Higgs boson. Jet candidates for which the distance is lower than the jet resolution parameter $\ycut$ are clustered together in the four-momentum recombination scheme. The algorithm stops when all pair of partons (or pseudo-jets) $(i,j)$ satisfy \mbox{$y^D_{ij}>\ycut$}. The $n$-jet rate, computed including terms up to order $\alpha_s^k$ in perturbative QCD, is denoted with $R^{(k)}_X(n, \ycut)$ and is defined as the fraction of events which has $n$ resolved jets for a given choice of jet resolution parameter $\ycut$:
\begin{align}
	R^{(k)}_X(n, \ycut) = \frac{\sigma^{(k)}_X(n,\ycut)}{\sigma^{(k)}_{X}}\,
	\label{eq:jetrate}
\end{align}
where $X$ represents the decay mode of the Higgs, i.e. $b\bar b$, $gg$ or the total sum, $\sigma^{(k)}_{X}$ is the inclusive hadronic cross section, and $\sigma^{(k)}_X(n,\ycut)$ is the exclusive cross section for the production of $n$ jets, with a jet-resolution parameter $\ycut$. 

Jet rates in hadronic Higgs decays have been computed up to order $\alpha_s^3$ with respect to the leading-order coupling in each channel in~\cite{Fox:2025cuz}. The calculation there is performed assuming a Higgs boson decaying at rest, hence in the centre-of-mass of the hadronic decay. The calculation we present here is performed in the laboratory frame with boosted Higgs (and $Z$) bosons and a hadronic cluster which is not at rest with respect to the centre-of-mass of the collision.  This modifies the fraction of $n$-jet events in both Higgs decay modes in a non-trivial way. In particular, 
because the leptonic recoil forces the hadronic radiation to be more collimated, 
particles in the decay products of the Higgs boson are more likely to be clustered into fewer jets, for a given value of $\ycut$. 
An immediate consequence of this is the presence of a non-vanishing one-jet rate at large $\ycut$ values, a situation encountered when all hadronic radiation is clustered in a single jet. This is not possible when the hadronic decays of a colour singlet are studied in its rest frame.

\begin{figure}
	\centering
	\includegraphics[width=0.48\linewidth]{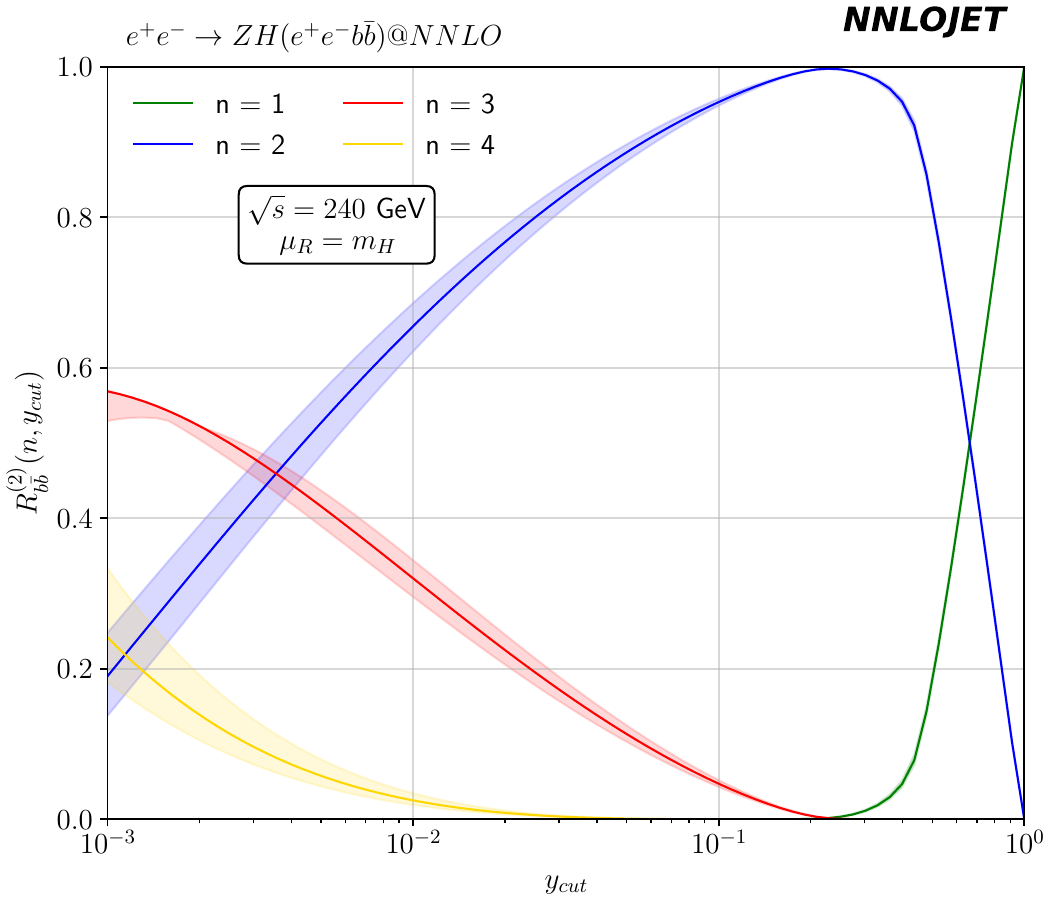}
	\includegraphics[width=0.48\linewidth]{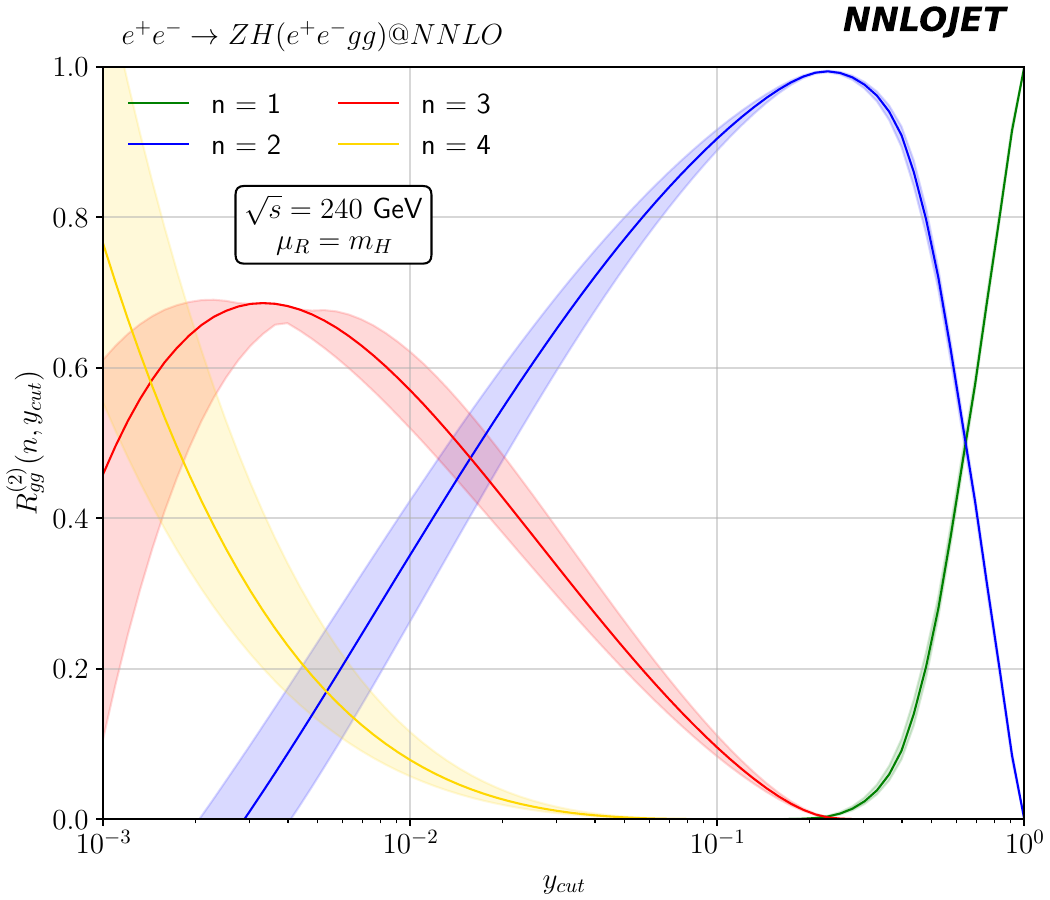}\\
	\includegraphics[width=0.48\linewidth]{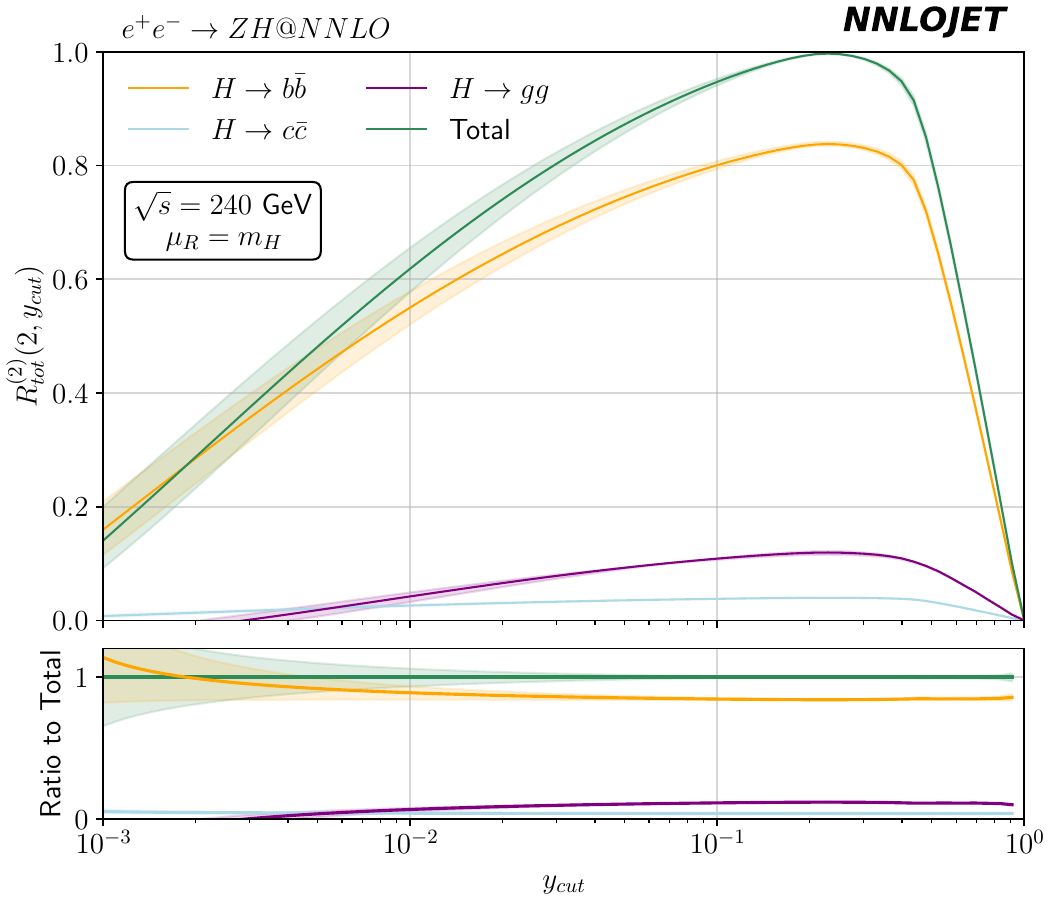}
	\includegraphics[width=0.48\linewidth]{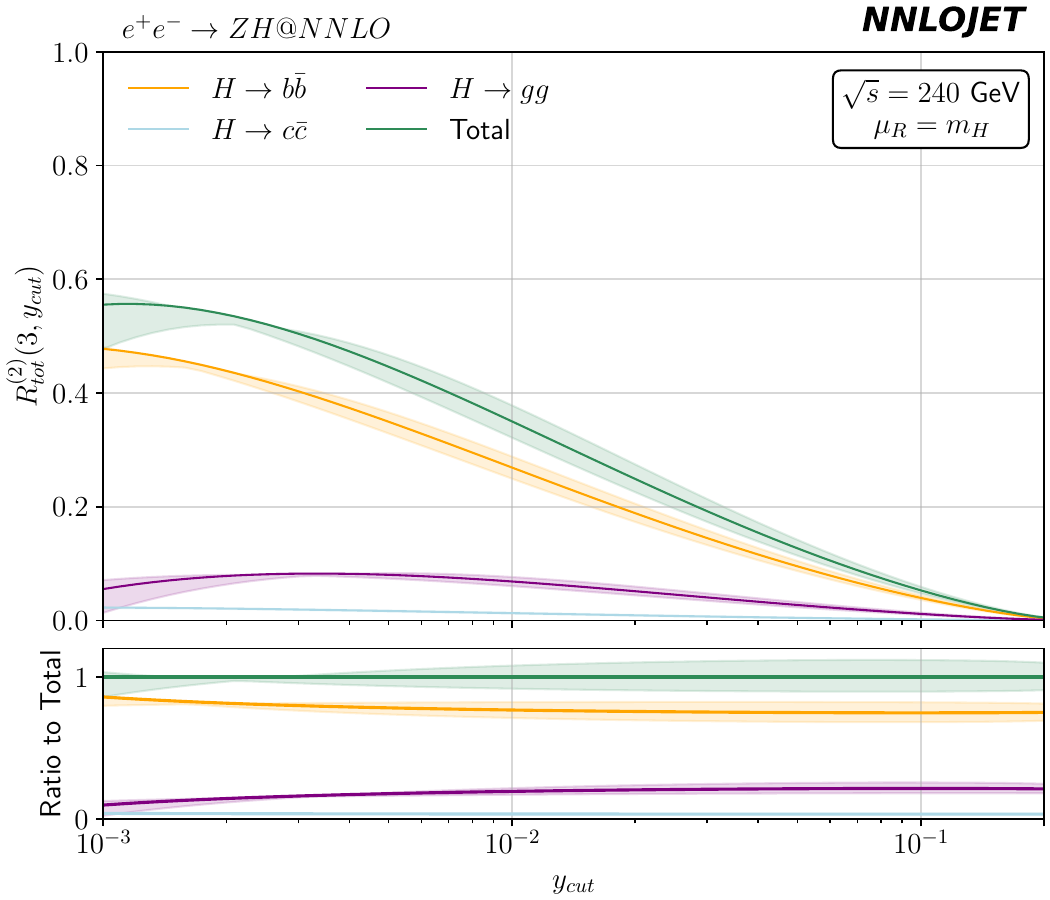}
	\caption{Top row: jet rates in the $H\rightarrow b\bar{b}$ (left) and $H\rightarrow gg$ (right) channels in $ZH$ production at lepton colliders. Bottom row: two-jet (left) and three-jet rate (right) for individual decay channel and the total sum, with the ratio to total shown in the lower frame.}
\label{fig:rates}
\end{figure}

Our calculation enables us to compute the one- and two-jet rates, denoted as $R^{(2)}_X(1, \ycut)$ and $R^{(2)}_X(2, \ycut)$, at NNLO, the three-jet rate $R^{(2)}_X(3, \ycut)$ at NLO and the four-jet rate $R^{(2)}_X(4, \ycut)$ at LO.  The renormalisation scale variation used to estimate theory uncertainties is performed in a correlated way in the numerator and denominator of~\eqref{eq:jetrate}. Our results are presented in Figure~\ref{fig:rates}, where we separately show the jet rates in the Yukawa and gluonic modes (top row), and the total sum of all decay channels for $R^{(2)}_X(2, \ycut)$ and $R^{(2)}_X(3, \ycut)$ (bottom row).

Analysing the general behaviour of the jet rates in both categories as a function of the jet resolution parameter $\ycut$, we notice, as expected, that
the large $\ycut$ regime is dominated by the one- and two-jet rates, with the former rising around $\ycut\approx0.2$ and rapidly overtaking the two-jet rate at $\ycut\approx 0.7$ and $\ycut\approx 0.6$ for the Yukawa and gluonic mode respectively.  The two-jet rate peaks at $\ycut\approx0.2$, where basically all events are classified as two-jet events.  For lower values of $\ycut$, the onset of the three- and four-jet rates is observed. As discussed in \cite{Fox:2025cuz} for a Higgs boson decaying at rest, the fraction of higher jet-multiplicity events at a given $\ycut$ value is larger in the gluonic case, due to the enhancement of emissions from gluon radiators. We observe that the intersection point between the two- and three-jet rate in the gluonic mode is located at values of $\ycut$ five times larger than in the Yukawa mode, respectively at $\ycut\approx0.015$ and $\ycut\approx0.003$. At even lower $\ycut$ values, the two-jet rate in the gluonic mode turns negative, as a consequence of the increased sensitivity to multiple soft emissions in this region, which invalidates the perturbative fixed-order description. Again, as observed in~\cite{Fox:2025cuz}, 
these effects also happen in the Yukawa decay mode, but they take place earlier, namely at larger $\ycut$ values, 
in the gluonic mode.

The theory uncertainty bands are in general larger for the $H\to gg$ decay channel, in particular for the three- and four-jet rate, which are here computed only at NLO and LO accuracy. In both decay modes, the scale uncertainty shrinks significantly at large values of $\ycut$, where multiple emissions are clustered together resulting in more inclusive and perturbatively stable results. 

The results shown in the bottom row of Figure~\ref{fig:rates} lead to analogous observations to those reported in~\cite{Fox:2025cuz}: the relative fraction of the different decay channels in the two-jet rate reflect those in the fully inclusive case ($85$\% for $H\to b\bar{b}$, $11$\% for $H\to gg$ and $4$\% for $H\to c\bar{c}$), even after the onset of the one-jet rate, while the fraction of gluonic events is enhanced in three-jet events at large $\ycut$.

Following the analysis of Figure~\ref{fig:rates} above, we can argue that by considering values of the jet resolution parameter  $\ycut$  in the window $0.01 \leq \ycut \leq 0.1$, it is possible to veto one-jet events (not particularly relevant for the observables considered in this paper), as well as probe higher jet-multiplicity events which can yield non-trivial kinematical effects, while avoiding the region where perturbative calculations become unreliable.  According to this fundamental observation, we choose an intermediate (logarithmically between $0.01$ and $0.1$) value of $\ycut$, namely $\ycut=0.03$, as our choice for the baseline predictions presented in the remainder of this section. Furthermore, we shall examine the $\ycut$ dependence for the sum over all decay channels at NNLO by comparing the baseline results with the ones obtained at $\ycut=0.01$ and $\ycut=0.1$.

\subsection{Jet Energies}
\label{sec:energies}

\begin{figure}[h]
  \centering
  \includegraphics[width=0.48\linewidth]{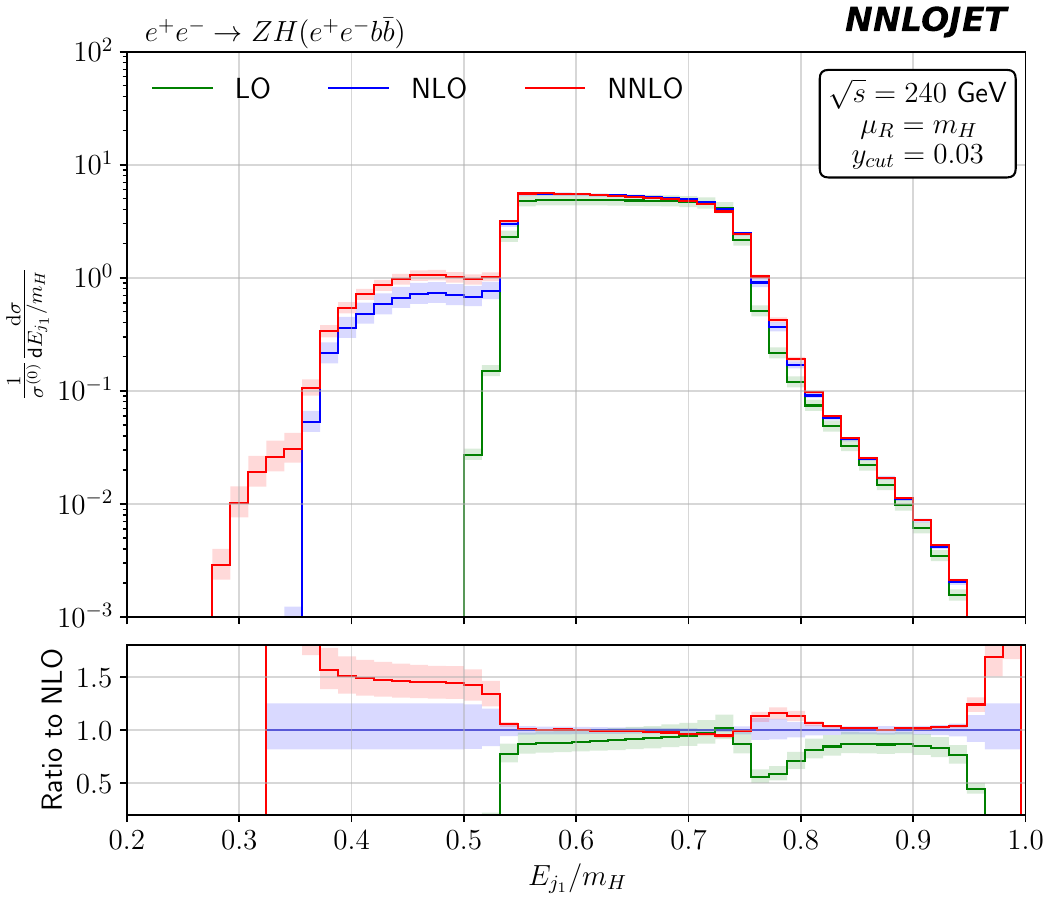}
  \includegraphics[width=0.48\textwidth]{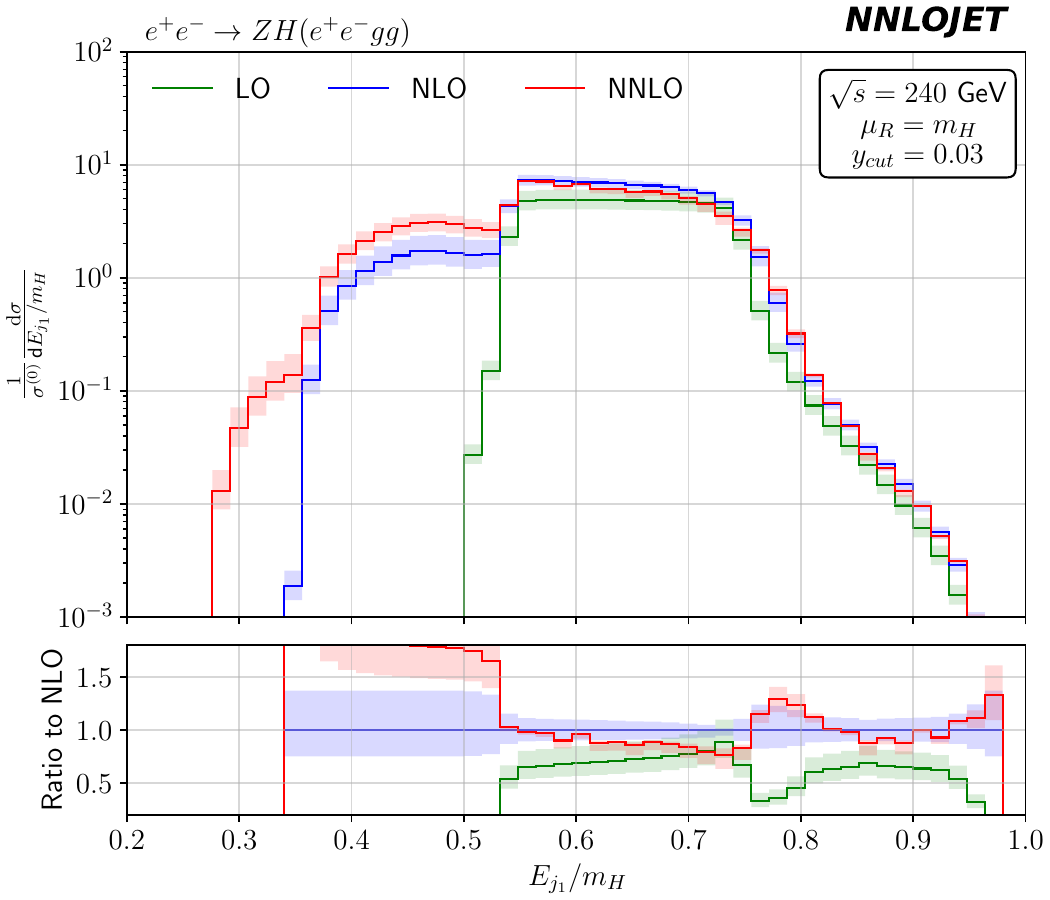}\\
  \includegraphics[width=0.48\linewidth]{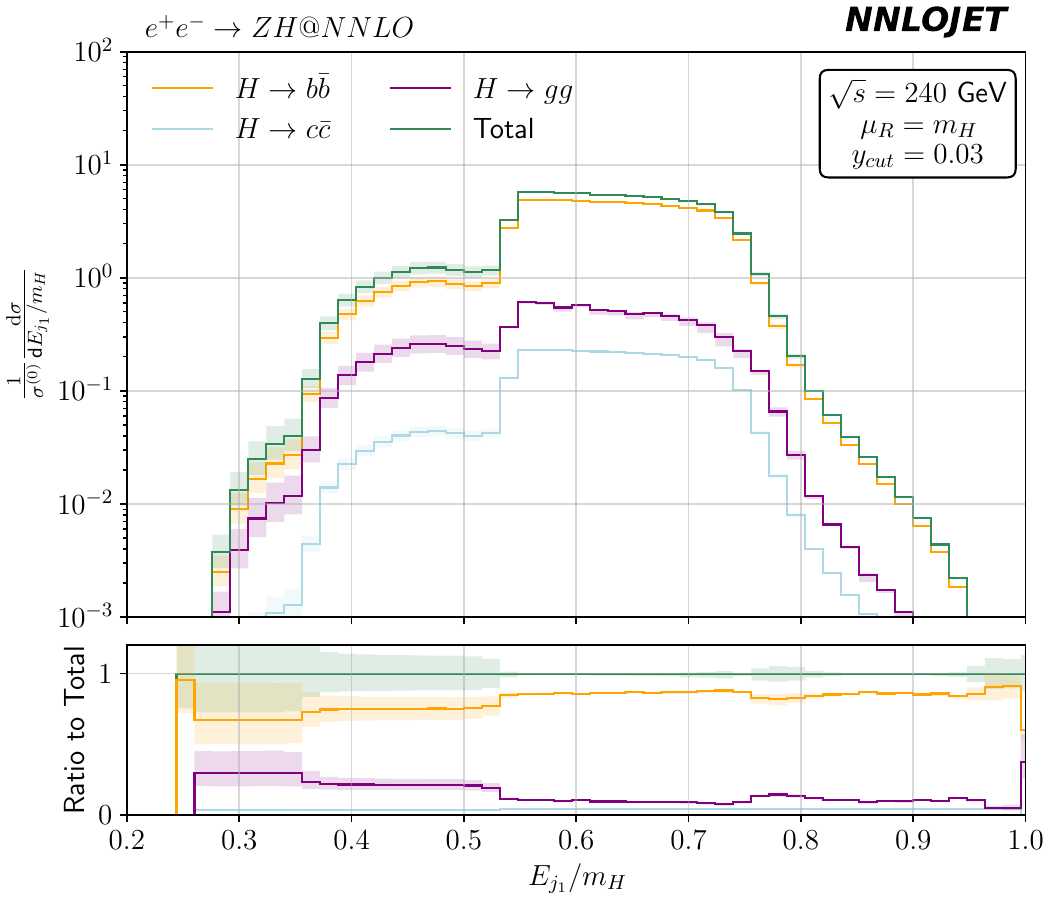}
  \includegraphics[width=0.48\textwidth]{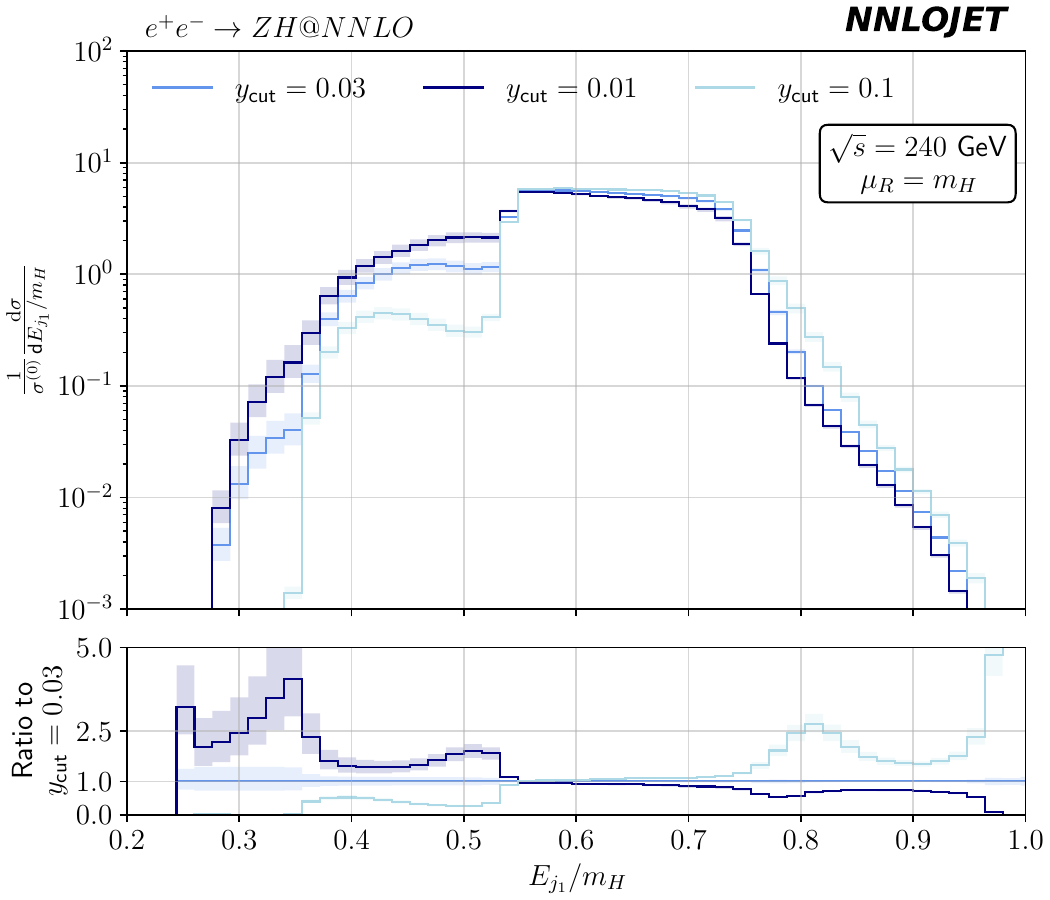}
  \caption{Energy distribution of the leading Durham-algorithm jets. In the top row: results for $\ycut = 0.03$ in the $H\to b\bar{b}$ (left) and the $H\to gg$ (right) channels at  LO (green), NLO (blue), and NNLO (red), with the ratio to NLO shown in the lower frame. Bottom left panel: comparison between the total sum of all decay channels (teal), $H\to b\bar{b}$ (yellow),  $H\to gg$ (purple), and $H\to c\bar{c}$ (light blue) at NNLO, with the ratio to the total shown in the lower frame. Bottom right panel: comparison between predictions for the sum over all decay channels at NNLO at $\ycut = 0.01$, $\ycut = 0.03$ and $\ycut = 0.1$, with the ratio to the $\ycut = 0.03$ results shown in the lower frame.}
  \label{fig:ej1}
\end{figure}
%

%Figure 4
\begin{figure}[h]
	\centering
	\includegraphics[width=0.48\linewidth]{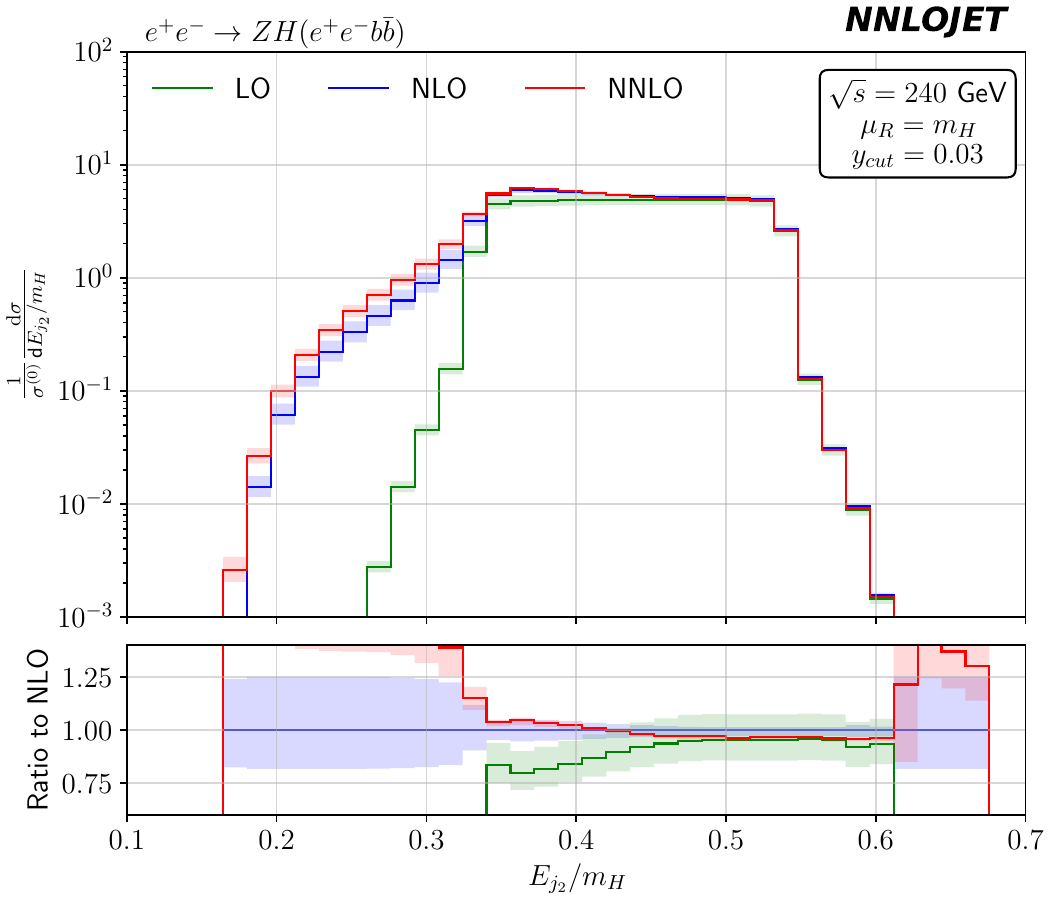}
	\includegraphics[width=0.48\textwidth]{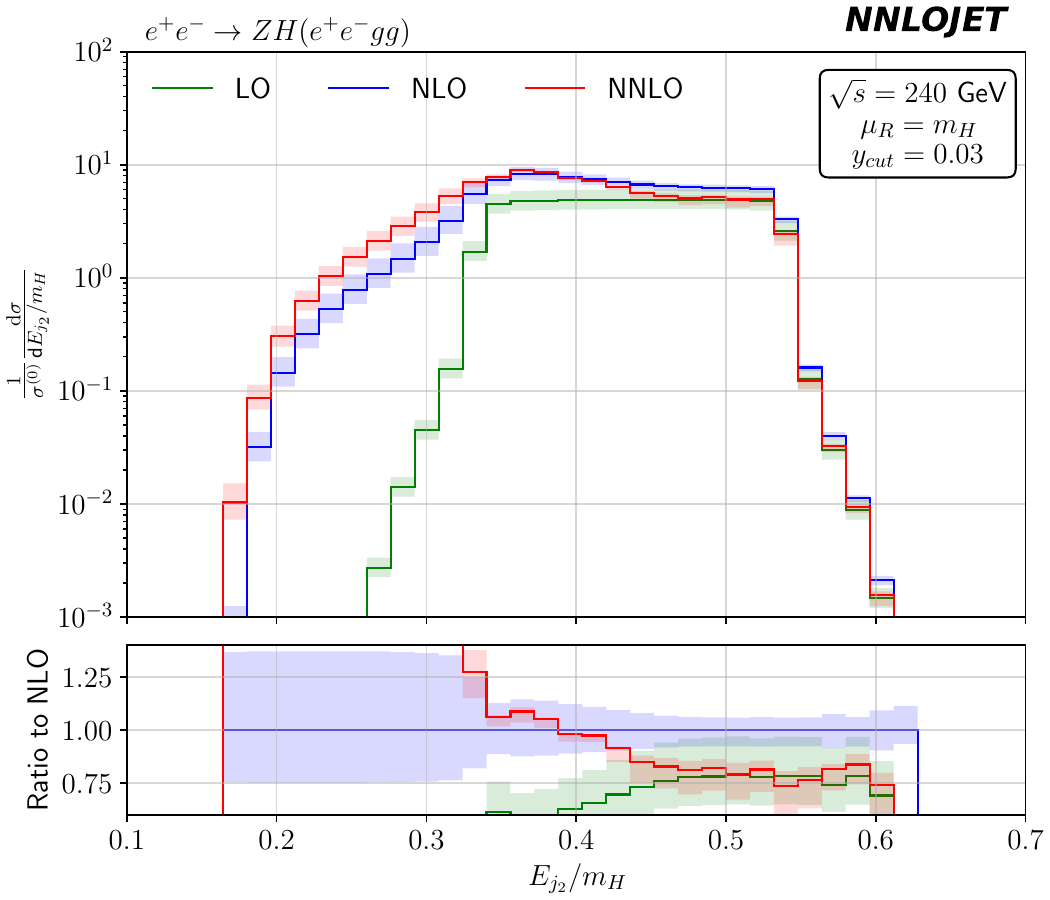}\\
	\includegraphics[width=0.48\linewidth]{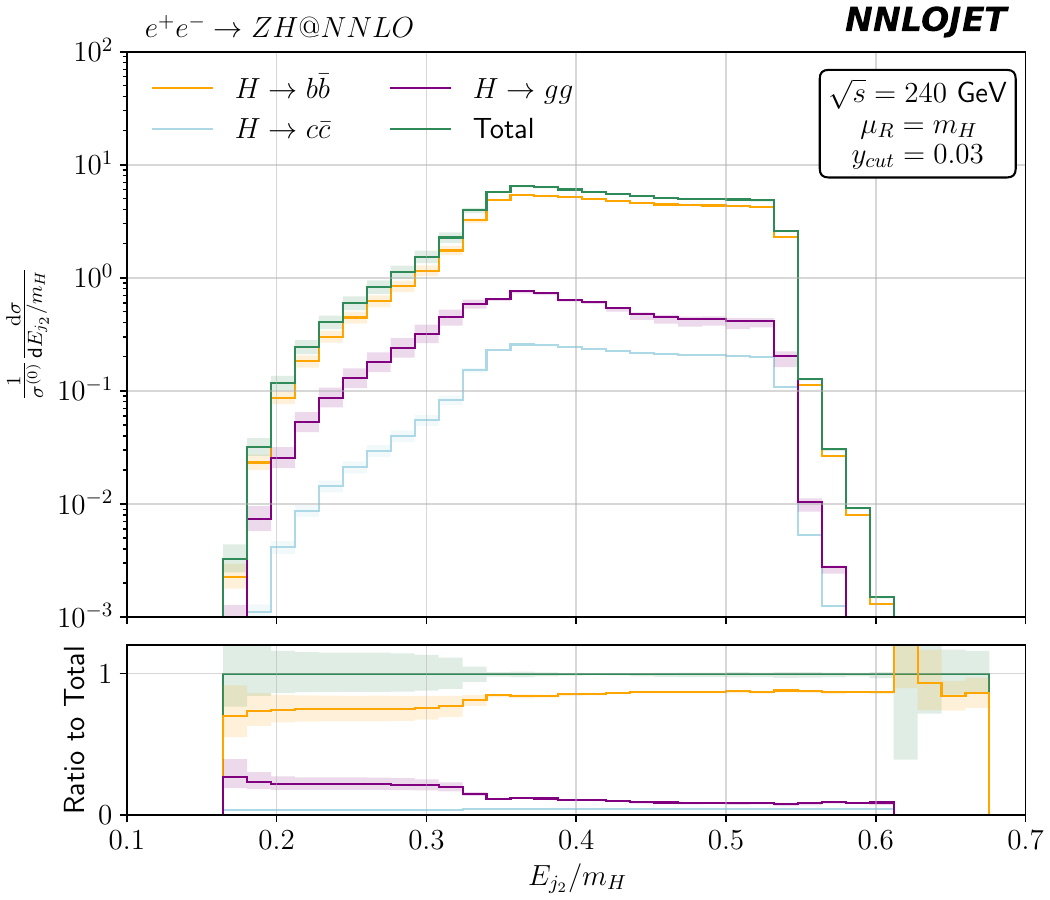}
	\includegraphics[width=0.48\textwidth]{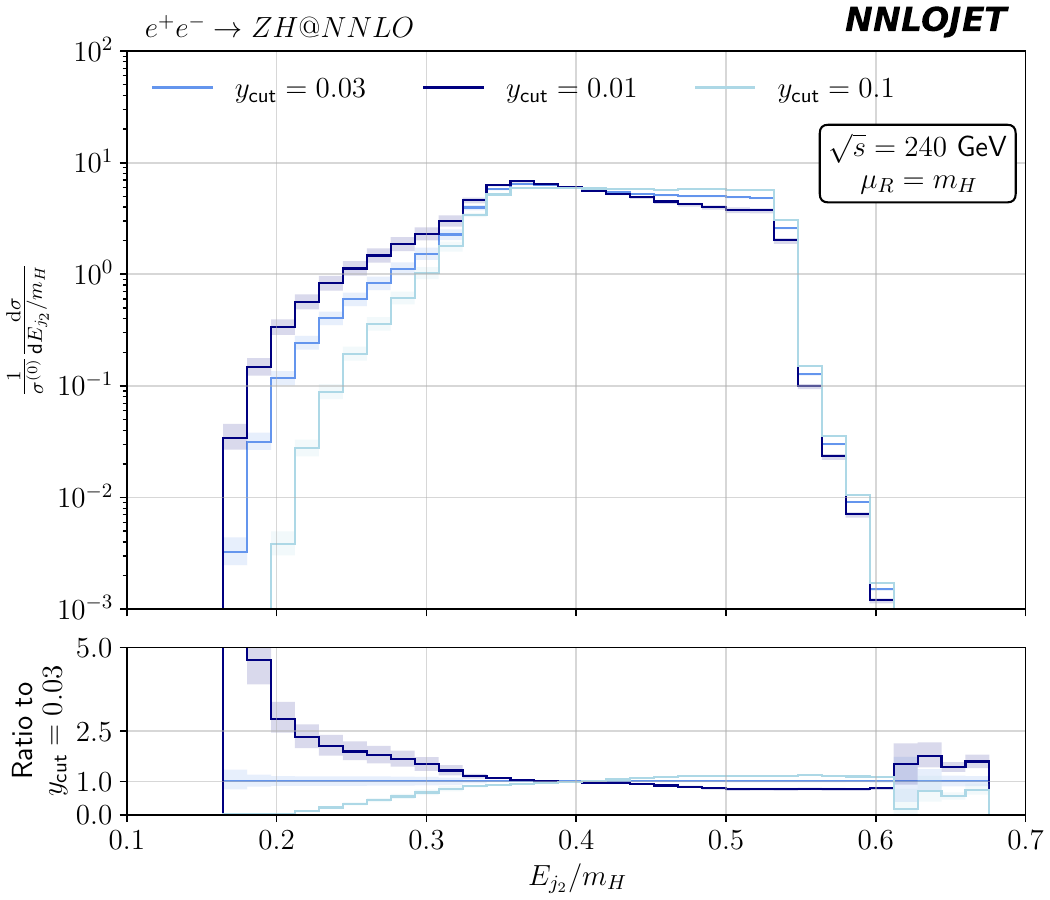}
	\caption{Energy distribution of the subleading Durham-algorithm jet. In the top row: results for $\ycut = 0.03$ in the $H\to b\bar{b}$ (left) and the $H\to gg$ (right) channels at  LO (green), NLO (blue), and NNLO (red), with the ratio to NLO shown in the lower frame. Bottom left panel: comparison between the total sum of all decay channels (teal), $H\to b\bar{b}$ (yellow),  $H\to gg$ (purple), and $H\to c\bar{c}$ (light blue) at NNLO, with the ratio to the total shown in the lower frame. Bottom right panel: comparison between predictions for the sum over all decay channels at NNLO at $\ycut = 0.01$, $\ycut = 0.03$ and $\ycut = 0.1$, with the ratio to the $\ycut = 0.03$ results shown in the lower frame.}
	\label{fig:ej2}
\end{figure}

% LO kinematics 
We start by examining distributions for the leading and subleading jet energy $E_{j_1}$ and $E_{j_2}$, presented in Figures~\ref{fig:ej1} and~\ref{fig:ej2} respectively. The distributions are computed requiring the presence of at least two resolved jets. This is clearly necessary to define $E_{j_2}$, but it is in principle not needed for $E_{j_1}$. One-jet events would give a contribution peaked at the energy of the Higgs boson, which we choose not to represent here. Our choice of the jet resolution parameter $\ycut=0.03$ anyway vetoes one-jet events, as discussed in Section~\ref{sec:jetrates}.  

A series of different kinematical regimes can be observed for the differential cross sections.
To analyse the results, we first consider the Born-level kinematics and assume on-shell $Z$ and Higgs bosons. In the laboratory frame, the Higgs boson and its decay products, the hadronic jets, are boosted. The energy of the Higgs boson in the rest frame ($m_H$) and in the laboratory frame are related by a boost factor $\gamma$ given by:

\begin{equation} 
\gamma = \dfrac{E^{\text{lab.}}_H}{m_H} = \dfrac{s + m_H^2 - m_Z^2}{2m_H\sqrt{s}}=1.082, 
\end{equation} 
leading to an energy of the Higgs boson in the laboratory frame of 
\begin{equation}
	E_H=\gamma m_H=135.347\,\,\,\gev.
\end{equation}
The boost factor $\gamma$ enters the definition of the energy for the leading and subleading jet as:
 \begin{equation}
 \dfrac{E^{\text{lab.}}_\pm}{m_H} = \frac{1}{2}\left(\gamma \pm \sqrt{\gamma^2-1} |\cos\theta|\right),
 \label{eq:boost_energy_jet} 
\end{equation} 
where $E^{\text{lab.}}_+$ is the energy of the leading jet ($j_1$) in the laboratory frame, while $E^{\text{lab.}}_-$ the one of the subleading jet
$(j_2)$, and $\theta$ is the angle between the direction of flight of the Higgs boson in the laboratory frame and the direction of the back-to-back decay products in the Higgs boson rest frame. 
Using \ref{eq:boost_energy_jet}, we can infer the minimum and maximum values of $E_{j_1}/m_H$ and $E_{j_2}/m_H$ at LO:
\begin{equation}
	\min\left(\dfrac{E_{j_1}}{m_H}\right)\approx0.54,\,\quad\max\left(\dfrac{E_{j_1}}{m_H}\right)\approx0.75,\,
\end{equation}
\begin{equation}
	\min\left(\dfrac{E_{j_2}}{m_H}\right)\approx0.33,\,\quad\max\left(\dfrac{E_{j_2}}{m_H}\right)\approx0.54,
\end{equation}
with the maximum value gained by the energy of the subleading jet 
corresponding to the minimum value attained by the energy of the leading jet, as expected. 
Importantly, the values found above can be read-off by scrutinising the LO results presented in the top row, for both Higgs decay modes, in Figures~\ref{fig:ej1} and~\ref{fig:ej2}. In particular, they define what can be called ``bulk region'' for the predictions. Before and after the bulk region, the leading order differential cross section is non-vanishing due to the off-shellness of the $Z$ and Higgs bosons, but it decreases sharply. 
In the bulk region, which is mostly populated by events with only two hard jets, we observe very good perturbative convergence for the $H\rightarrow b\bar{b}$ decay mode for both observables, with small higher-order corrections and percent-level left-over theory uncertainties at NNLO. On the contrary, NLO and NNLO corrections are much more sizeable in the $H\rightarrow gg$ mode, with approximately $\pm10$\% residual theory uncertainties at NNLO. Moreover, poor perturbative convergence is observed, with the NLO scale variation bands not fully containing the NNLO results. This can be explained by the fact that real emissions are enhanced in the gluonic decay mode, leading to a high sensitivity to radiative corrections in the bulk region, dominated by two-jet exclusive events.

To the left of the bulk region, for both Higgs decay modes, we can clearly identify the opening of three- and four-jet contributions beyond LO. These configurations populate a previously forbidden region of the phase space and extend the kinematical range in these energy distributions. Formally having LO and NLO accuracy, the corresponding  
scale uncertainty bands are larger here than those encountered for the two-jet contributions in the bulk region, and more pronounced in the gluonic decay mode. This is in close analogy to what was observed for the jet rates in the top 
row of Figure~\ref{fig:rates}. To the right of the bulk region, the leading and subleading energy distributions decrease sharply at each perturbative order, with larger corrections and residual theory uncertainties than in the bulk region for both Higgs decay categories.   

As shown in the bottom left frames of Figures~\ref{fig:ej1} and~\ref{fig:ej2}, the relative fractions of the different decay channels in the bulk region and above it are roughly constant and coincide with the inclusive branching ratios. By contrast, in the left tail, the fraction of gluonic events is enhanced, as expected from the fact that in this region higher jet-multiplicities are relevant. In the leading-jet energy distribution shown in Figure~\ref{fig:ej1}, it is possible to observe a first enhancement where three-jet events dominate, followed by a second one due to four-jet events. 

In the bottom right frames of Figures~\ref{fig:ej1} and~\ref{fig:ej2}, we compare our baseline predictions at $\ycut=0.03$ for the total sum over decay channels to results obtained at $\ycut=0.01$ and $\ycut=0.1$. For both observables, we observe mild differences in the bulk region, populated by exclusive two-jet events, with larger $\ycut$ values giving larger cross sections, as expected. A similar behaviour continues above the bulk region. To the left of the bulk region, the hierarchy is inverted, with smaller $\ycut$ values yielding larger contributions, due to the enhanced relevance of three- and four-jet events. 

\subsection{Angular observables}
\label{sec:angles}

In this section we present results for angular observables, in particular for the angle between the leading and the subleading jets $\cos\theta_{j_1j_2}$ and the angle between the negatively charged lepton and its closest jet $\cos\theta_{l^-j_{cl.}}$. 
\begin{figure}[h]
	\centering
	\includegraphics[width=0.48\linewidth]{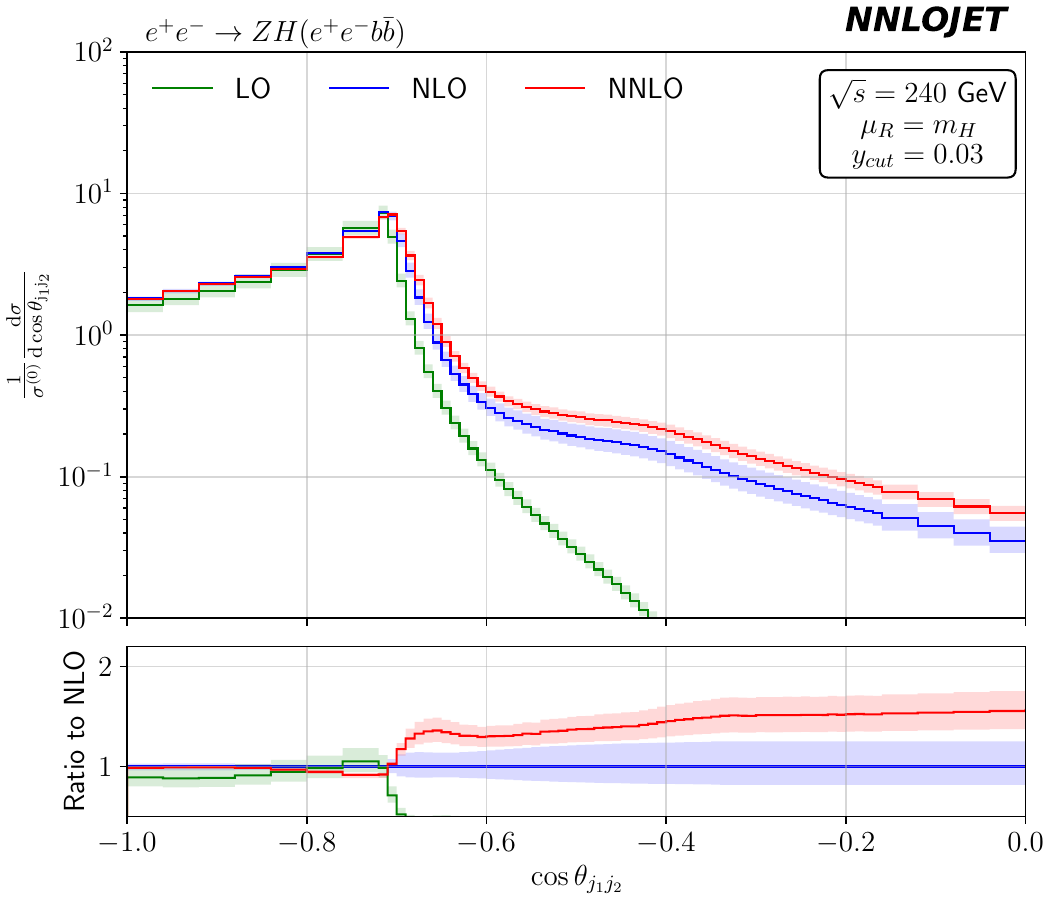}
	\includegraphics[width=0.48\textwidth]{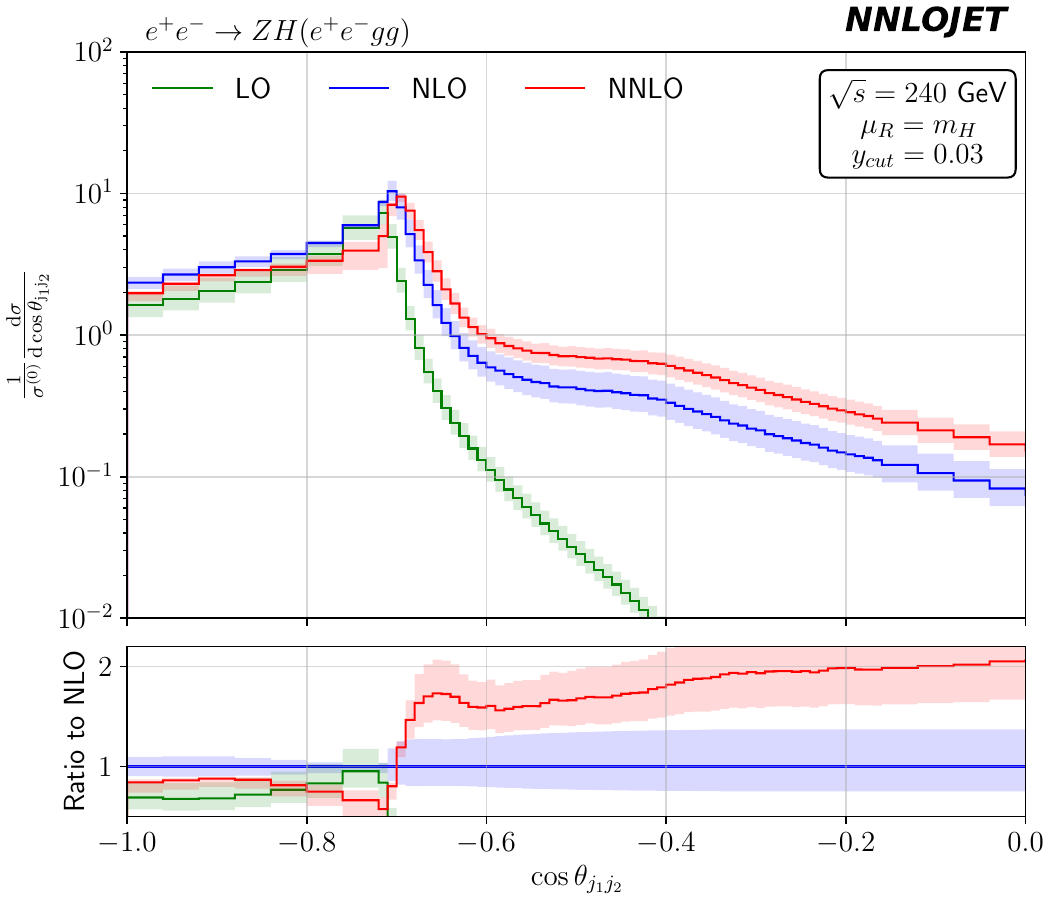}\\
	\includegraphics[width=0.48\linewidth]{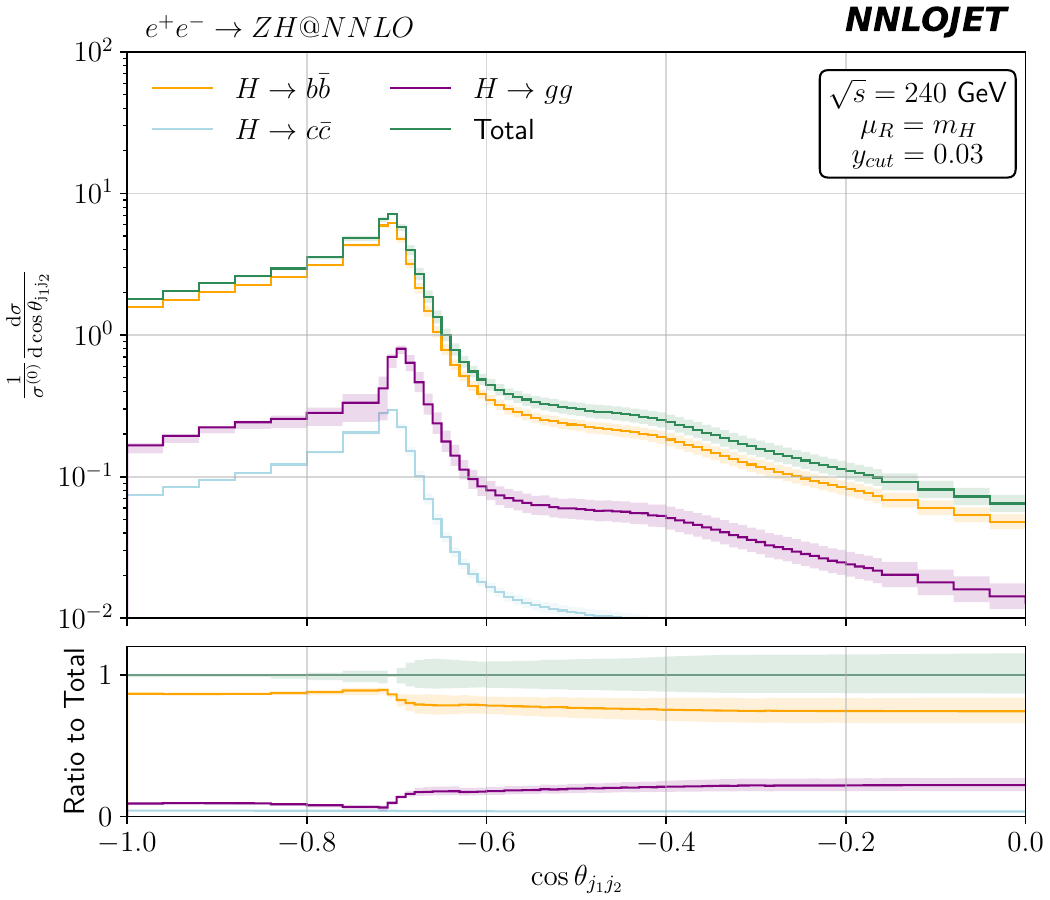}
	\includegraphics[width=0.48\textwidth]{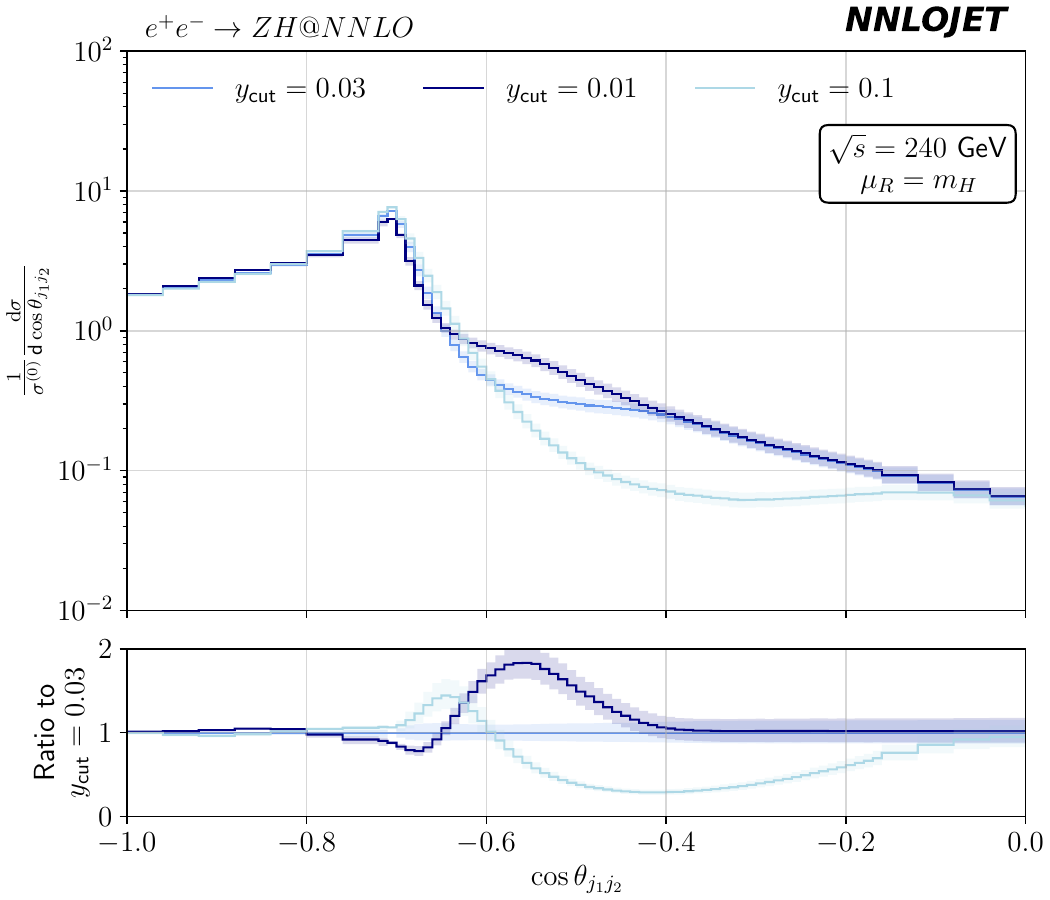}
	\caption{Distribution of the angle between leading and subleading Durham-algorithm jet. In the top row: results for $\ycut = 0.03$ in the $H\to b\bar{b}$ (left) and the $H\to gg$ (right) channels at  LO (green), NLO (blue), and NNLO (red), with the ratio to NLO shown in the lower frame. Bottom left panel: comparison between the total sum of all decay channels (teal), $H\to b\bar{b}$ (yellow),  $H\to gg$ (purple), and $H\to c\bar{c}$ (light blue) at NNLO, with the ratio to the total shown in the lower frame. Bottom right panel: comparison between predictions for the sum over all decay channels at NNLO at $\ycut = 0.01$, $\ycut = 0.03$ and $\ycut = 0.1$, with the ratio to the $\ycut = 0.03$ results shown in the lower frame.}
	\label{fig:costh_j12}
\end{figure}

We first consider differential distributions in $\cos\theta_{j_1j_2}$, given in Figure~\ref{fig:costh_j12}. Again, it is instructive to determine the opening angle between the two jets in the LO kinematics, assuming on-shell bosons. We obtain:
\begin{equation}
	\cos\theta_{j_1j_2}=-\dfrac{1-2\beta^2+\beta^2\cos^2\theta}{1-\beta^2\cos^2\theta},
	\label{costheta12}
\end{equation}
where $\beta$ is the velocity of the Higgs boson in the laboratory frame:
\begin{equation}
	\beta=\sqrt{1-\dfrac{1}{\gamma^2}}\approx0.381,
\end{equation}
and, as in Section~\ref{sec:energies}, $\theta$ is the angle between the direction of flight of the Higgs boson in the laboratory frame and the direction of the back-to-back decay products in the Higgs boson rest frame. The expression in~\eqref{costheta12} reaches the minimum value $-1$ for $\theta=n\pi$, while its maximum is at $\theta=\left(n+\frac{1}{2}\right)\pi$ and is given by:
\begin{equation}
	\max\left(\cos\theta_{j_1j_2}\right)=-1+2\beta^2\approx-0.710,
\end{equation}
corresponding to an opening angle of approximately $ 0.75\pi$. With this, we can explain why the LO distributions in the top row of Figure~\ref{fig:costh_j12} exhibit a sharp decay after the maximum value of $\cos\theta_{j_1j_2}$, corresponding to the position of the peak, is reached. We can then define the bulk region for this angular observable to be $-1\leq\cos\theta_{j_1j_2}\lesssim-0.710$. This window includes the peak and the region to the left of it, which corresponds to a larger angular separation between the jets.  

Similarly to the observations made for the leading- and subleading-jet energy, in the bulk region higher-order corrections and the corresponding residual theory uncertainties are small for the $H\to b\bar{b}$ decay mode, while poorer perturbative convergence is seen in the $H\to gg$ channel. Here, NNLO corrections amount to $-15$\% of the NLO results to the left of the peak and lie outside of the NLO uncertainty bands. We again interpret this in terms of the higher sensitivity to additional emissions  for the gluonic channel. 

Radiative corrections allow to populate the phase space to the right of the bulk region, corresponding to a smaller opening angle between the two jets. Here three- and four-jet events are relevant. The formal perturbative accuracy is then reduced, hence higher-order corrections and residual scale uncertainties are very sizeable. Along with what was observed so far, the gluonic mode exhibits the largest corrections, with NNLO results in the right tail of the distributions being $100\%$ larger than NLO ones. This is also reflected in the bottom left frame of Figure~\ref{fig:costh_j12} by the increased fraction of gluonic decays to the right of the bulk region, where higher jet-multiplicity events dominate. Sudakov shoulders appear at the peak of the distribution, as well as at the point where the LO predictions vanish ($\cos\theta_{j_1j_2}\approx-0.4$) and a change of slope is observed starting from NLO.

\begin{figure}[h]
	\centering
	\includegraphics[width=0.48\linewidth]{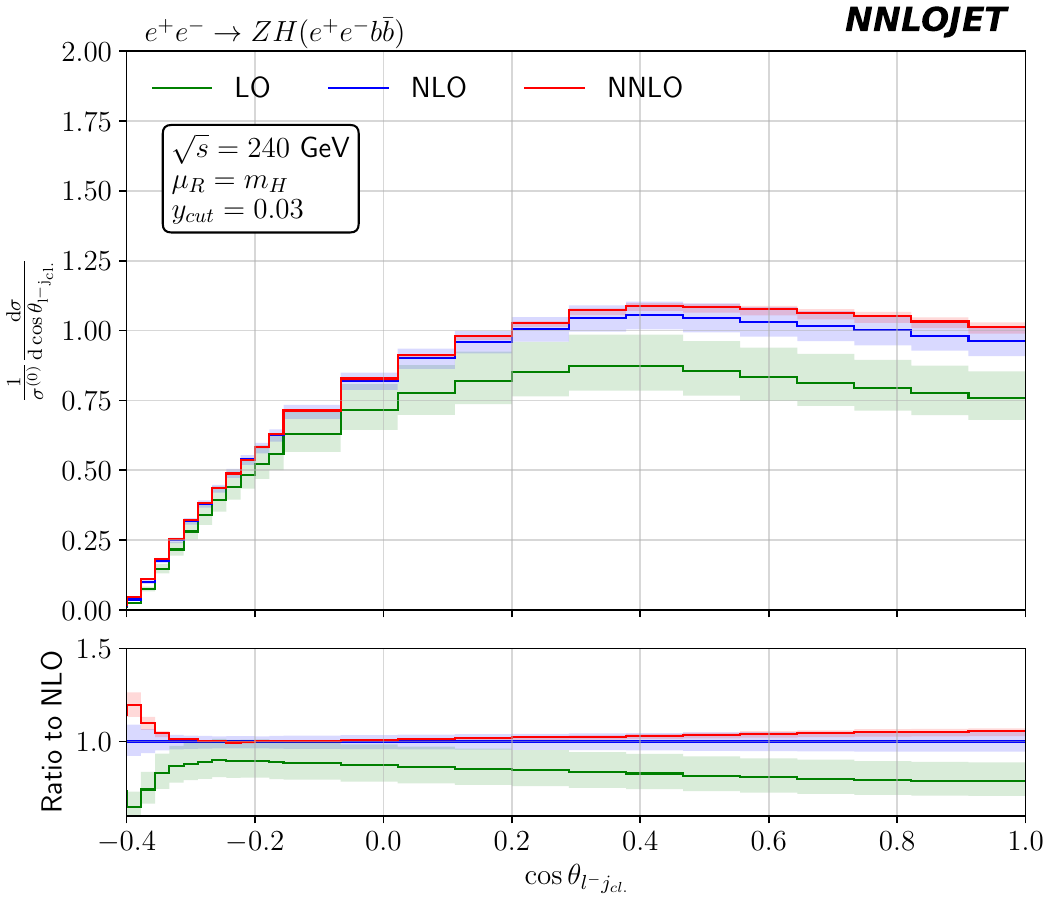}
	\includegraphics[width=0.48\textwidth]{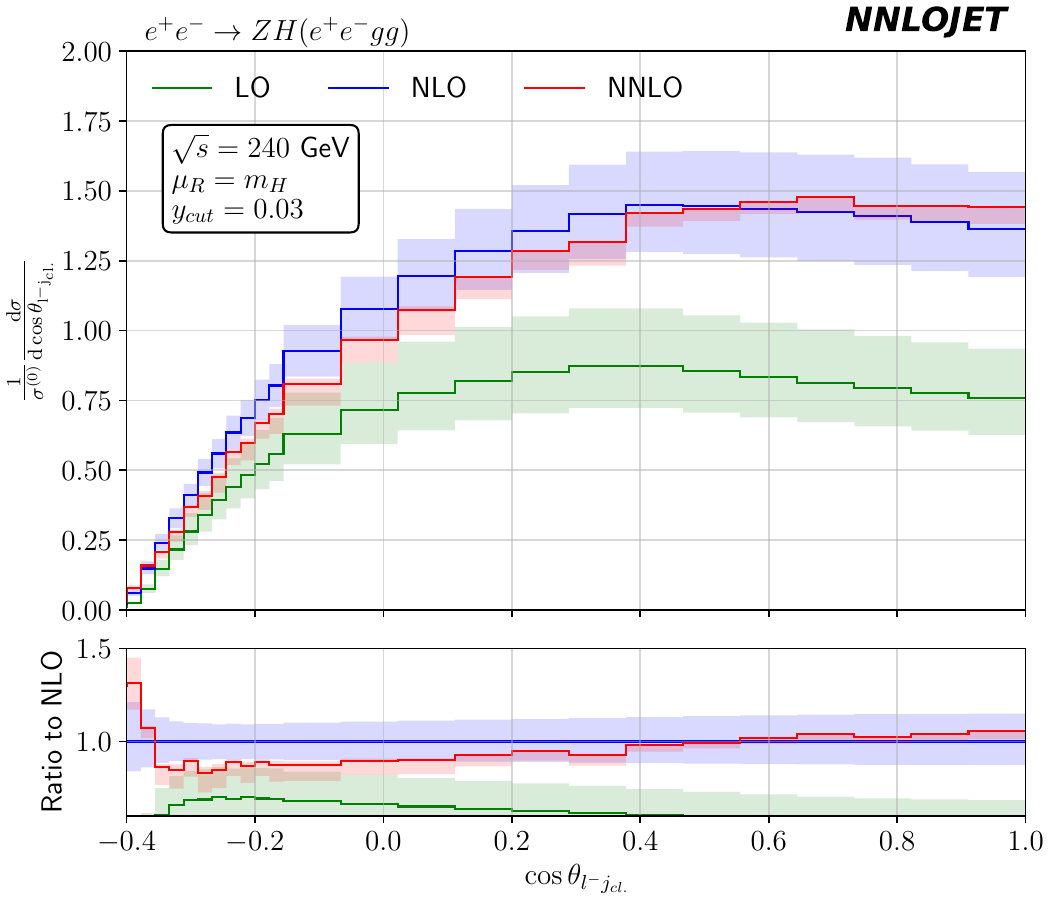}\\
	\includegraphics[width=0.48\linewidth]{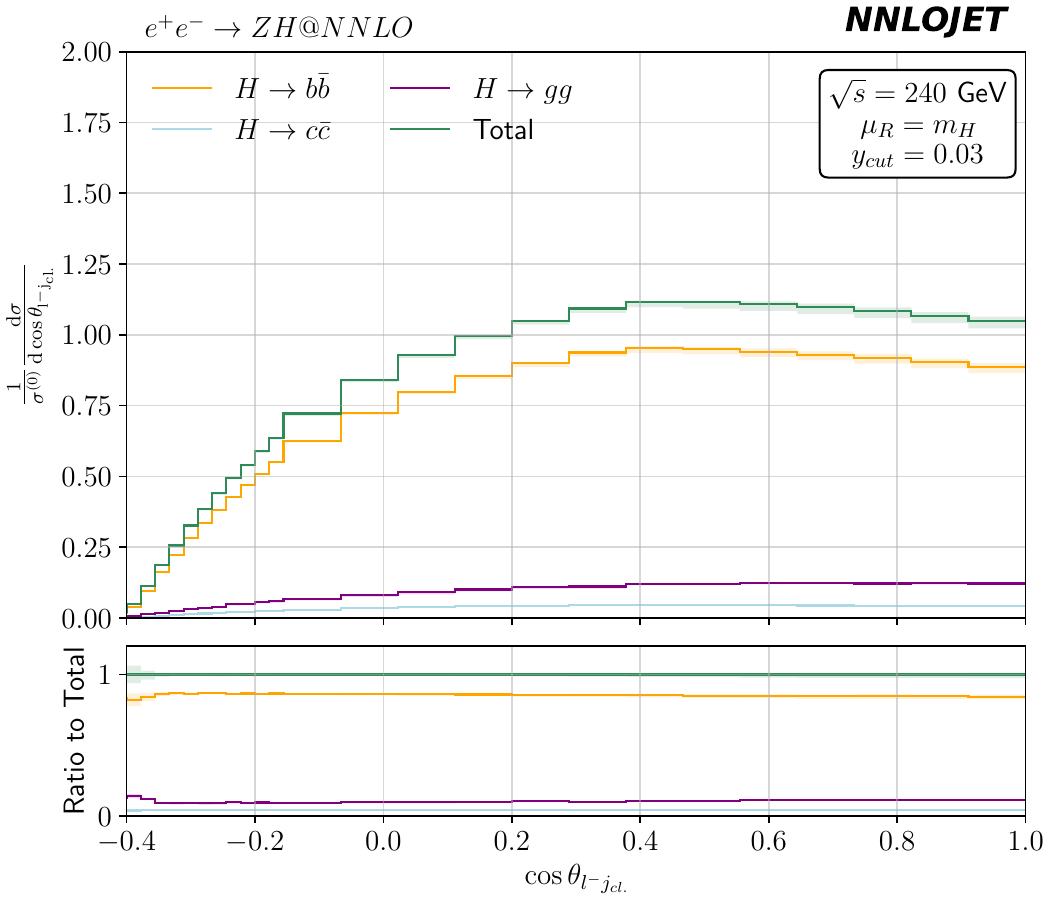}
	\includegraphics[width=0.48\textwidth]{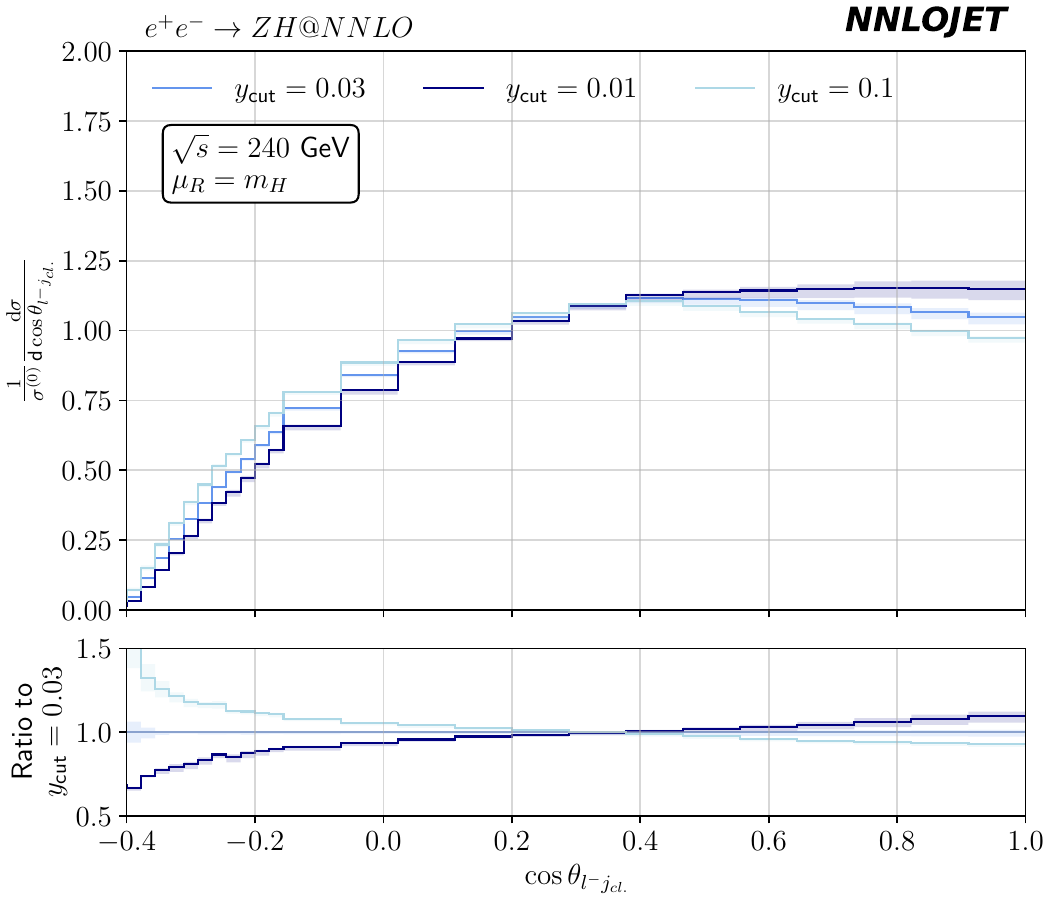}
	\caption{Distribution of the angle between the negatively charged lepton and its closest Durham-algorithm jet. In the top row: results for $\ycut = 0.03$ in the $H\to b\bar{b}$ (left) and the $H\to gg$ (right) channels at  LO (green), NLO (blue), and NNLO (red), with the ratio to NLO shown in the lower frame. Bottom left panel: comparison between the total sum of all decay channels (teal), $H\to b\bar{b}$ (yellow),  $H\to gg$ (purple), and $H\to c\bar{c}$ (light blue) at NNLO, with the ratio to the total shown in the lower frame. Bottom right panel: comparison between predictions for the sum over all decay channels at NNLO at $\ycut = 0.01$, $\ycut = 0.03$ and $\ycut = 0.1$, with the ratio to the $\ycut = 0.03$ results shown in the lower frame.}
	\label{fig:costh_lmj}
\end{figure}

In the bottom right frame of Figure~\ref{fig:costh_j12}, we study the effect of different $\ycut$ choices. We notice negligible differences in the bulk region to the left of the peak, which is dominated by two-jet events, while the prediction deviate significantly from each other at the peak and to the right of it. No clear hierarchy is observed in the window $-0.710\lesssim\cos\theta_{j_1j_2}\lesssim-0.6$, while for larger values of $\cos\theta_{j_1j_2}$, lower $\ycut$ choices lead to a larger cross section, due to the enhancement of events with three or more resolved jets. Predictions at $\ycut=0.01$ and $\ycut=0.03$ become identical around $\cos\theta_{j_1j_2}\approx-0.4$, indicating that there are no differences in the reconstruction of the two leading jets with these resolution parameter choices beyond this point. Similarly, the $\ycut=0.1$ predictions merge too at the right end of the shown spectrum.

We then consider the distribution related to the opening angle between the negatively charged lepton and its closest jet, presented in Figure~\ref{fig:costh_lmj}. This observable probes the interplay between the leptonic and the hadronic cluster. We notice however, that, given the scalar nature of the Higgs boson, the observable is not affected by spin correlations between the decay products of the two bosons.

It is again possible to define a bulk region for this observable considering LO kinematics and on-shell bosons. Contrary to the observables considered before, which were defined according to the sole kinematics of the hadronic final state, an analytical derivation of $\cos\theta_{l^-j_{cl.}}$ demands a parametrisation of the complete four-particle phase space and is therefore more complicated. We then resort to a numerical calculation to identify the boundary of the bulk region, and we find that $-0.4\lesssim\cos\theta_{l^-j_{cl.}}\leq 1$. The lower bound corresponds to a maximum angle of $\approx0.63\pi$. Off-shellness effects give a non-zero, although rapidly vanishing, contribution also for lower values of $\cos\theta_{l^-j_{cl.}}$. Since in this case higher jet-multiplicity final states do not extend the phase space accessible at LO, we choose to restrict the $x$-axis range of the plots in Figure~\ref{fig:costh_lmj} to precisely the bulk region identified above. 

The cross section increases from the lower to larger values of $\cos\theta_{l^-j_{cl.}}$, reaching its maximum around $\cos\theta_{l^-j_{cl.}}\approx0.4$, after which the distribution remains roughly constant with a mildly decreasing trend. In the considered range, NLO corrections are positive, relatively flat and sizeable for both decay modes, reaching up to $30$\% and $80$\% in the Yukawa and gluonic channel respectively. NNLO corrections are positive and small ($<5$\%) for the $H\to b\bar{b}$ decay mode, and good perturbative convergence is observed here. On the other hand, they are negative (apart from the rightmost bins) for the gluonic mode and reach up to $-10$\% of the NLO around $\cos\theta_{l^-j_{cl.}}\approx-0.3$. For $\cos\theta_{l^-j_{cl.}}\lesssim0.2$, the NLO scale variation bands fail to capture the NNLO corrections and the residual uncertainties amount to $\pm10$\%, similarly to what was found for other observables.

The relative fraction of each decay channel is constant across the displayed kinematical range, once again mirroring the inclusive case. For  of $0\lesssim\cos\theta_{l^-j_{cl.}}\lesssim0.5$, no relevant differences are observed varying $\ycut$, indicating that the angular separation between the leptons and their closest jet is not sensitive to the details of the reconstruction algorithm. For larger values of $\cos\theta_{l^-j_{cl.}}$, namely smaller opening angles, smaller $\ycut$ values provide a larger cross section. Indeed, when a multi-jet event with a jet close in angle to the negatively charged lepton is re-clustered with a larger $\ycut$ value, if the closest jet was not the leading one in energy, the angle $\theta_{l^-j_{cl.}}$ likely increases, hence $\cos\theta_{l^-j_{cl.}}$ decreases. Instead, for lower values of $\cos\theta_{l^-j_{cl.}}$, larger $\ycut$ choices lead to increasingly larger cross sections: if more jets are resolved, it's more likely that the opening angle between the negatively-charged lepton and its closest jet is smaller than the maximum one.

\section{Conclusions}
\label{sec:conclusions}
	In this paper we present NNLO-accurate predictions for the associated production of a $Z$ and a Higgs boson at lepton colliders, with the $Z$ boson decaying leptonically and the Higgs boson decaying hadronically into two or more hard jets. This process constitutes the dominant production channel for Higgs bosons at future electron-positron colliders, offering opportunities for high-precision studies of the Higgs boson properties and for the measurement of its couplings to other Standard Model particles. 
	
	We consider the dominant hadronic decay channels of a Higgs boson to bottom or charm quarks via Yukawa interaction and to gluons via an effective vertex obtained in the infinite top-quark mass limit. We study differential distributions in a series of jet observables, highlighting differences in the behaviour of the perturbative corrections in the Yukawa and gluonic modes, as well as investigating the impact of different jet resolution parameter choices for the Durham (or $k_T$) jet algorithm. We also provide phenomenological predictions for the sum of all decay channels, identifying regions where the relative fraction of different decay modes deviate significantly from the inclusive case. 
	
	We largely confirm what has been observed as a result of the perturbative computations for the hadronic decays of a Higgs boson in its rest frame~\cite{Fox:2025cuz,Fox:2025qmp}. In particular, the gluonic decay mode is more sensitive to radiative corrections, with the contribution of higher jet-multiplicity final-states being enhanced with respect to the Yukawa decay channel. These effects are clearly visible in all the observables we considered. In relation to this, the gluonic mode typically involves larger higher-order corrections and theoretical uncertainties, poorer perturbative convergence and an earlier breakdown of the perturbative approach, due to the increased sensitivity to multiple soft emissions from hard gluon radiators. 

Since in this work we consider the complete production process $e^+e^-\to ZH$, 
in our results we investigate the impact of the fact that the Higgs and $Z$ boson together with their decay products are boosted, and observed that this	induces non-trivial modifications for the differential distributions.
To interpret these observations, and in particular the onset of qualitatively different regimes in our results, we find convenient to refer to the kinematics of the underlying Born-level process. 
Because the initial state is known completely and the leptonic final state can be fully reconstructed experimentally, one can derive simple relations between the physical parameters and the observables, which are visible in the distributions. 
	
Our results complement a series of recent precision calculations of hadronic Higgs boson decays~\cite{Gao:2019mlt,Gao:2020vyx,Fox:2025cuz,Fox:2025qmp,Gehrmann-DeRidder:2023uld,Gehrmann-DeRidder:2024avt,Fox:2025txz,Alioli:2020fzf}, aimed at exposing relevant differences between the various decay modes.  Possible future directions include reaching NNLO accuracy in the three-jet region as in~\cite{Fox:2025qmp},
and considering an hadronically-decaying $Z$ boson to investigate the interplay between two separate hadronic clusters.

\acknowledgments
We would like to thank Christian Preuss and Mattia Pozzoli for their contributions at a preliminary stage of this project. We are also grateful to Elliot Fox, Thomas Gehrmann and Nigel Glover for useful discussions and feedback on the manuscript. AG and SC acknowledge the support of the Swiss National Science Foundation (SNF) under contract 200021-231259 and of the Swiss National Supercomputing Centre (CSCS) under project ID ETH5f. The work of SC has been partially supported by the Italian Ministry of Universities and Research (MUR) under the FIS grant (CUP: D53C24005480001, FLAME).
MM is supported by a Royal Society Newton International Fellowship (NIF/R1/232539).

\bibliography{bibliography.bib}

\end{document}